\documentclass[12pt,letterpaper,usenames,dvipsnames]{article}
\usepackage{jheppub}\usepackage{amsmath,amssymb,amsfonts,amsthm,amscd}
\usepackage{bm} 
\usepackage{bbm} 
\usepackage{mathrsfs} 
\usepackage{lscape} 
\usepackage{subcaption} 
\usepackage{tcolorbox}
\usepackage{multirow} 
\usepackage{colortbl} 
\usepackage{array,hhline,arydshln,multirow}
\usepackage{floatrow}
\usepackage[export]{adjustbox}
\usepackage{rotating} 

\newcommand{\bea}{\begin{eqnarray}}
\newcommand{\eea}{\end{eqnarray}}
\newcommand{\be}{\begin{equation}}
\newcommand{\ee}{\end{equation}}

\usepackage{xstring}
\usepackage{tikz} 
\usepackage{booktabs}
\usetikzlibrary{decorations.pathreplacing} 
\usetikzlibrary{decorations.pathmorphing} 
\usetikzlibrary{decorations.markings} 
\usetikzlibrary{arrows} 
\usetikzlibrary{shapes} 
\usetikzlibrary{matrix} 
\usetikzlibrary{positioning} 
\usetikzlibrary{calc}

\tikzstyle{brane}=[draw]
\tikzset{D7/.style={circle, draw=black, inner sep=0pt, fill=white, minimum size=3mm}}
\tikzset{hasse/.style={circle, fill,inner sep=2pt}}
\tikzset{flavor/.style={regular polygon,regular polygon sides=4,inner sep=2.5pt, draw}}
\tikzset{gauge/.style={circle, draw,inner sep=2.5pt}}
\tikzset{biggauge/.style={circle, inner sep=4pt, fill=gray}}
\tikzset{gaugeb/.style={circle, draw,fill=black,inner sep=2.5pt}}
\tikzset{gauger/.style={circle, draw,fill=red,inner sep=2.5pt}}
\tikzset{bd/.style={circle, draw=black, inner sep=0pt, fill=black, minimum size=2mm}}
\tikzset{wd/.style={circle, draw=black, inner sep=0pt, fill=white, minimum size=2mm}}
\tikzset{Dynkin/.style={circle, draw=black, inner sep=0pt, fill=white, minimum size=2mm}}
\tikzstyle{ligne}=[draw, thick] 
\tikzstyle{gridline}=[draw, gray] 
\tikzset{doublearrow/.style={ draw=black!75, color=black!75, thick, double distance=3pt, }}

\allowdisplaybreaks[2]  
\usepackage{pifont,dsfont}
\usepackage{bm} 

\usepackage{xcolor}

\definecolor{amaranth}{rgb}{0.9, 0.17, 0.31}
\definecolor{coolblack}{rgb}{0.0, 0.18, 0.39}
\definecolor{gold(web)(golden)}{rgb}{1.0, 0.84, 0.0}
\definecolor{deepcarmine}{rgb}{0.66, 0.13, 0.24}

\def\bM{\begin{matrix}}
\def\eM{\end{matrix}}
\newcommand{\bpm}{\begin{pmatrix}}
\newcommand{\epm}{\end{pmatrix}}
\newcommand{\bsm}{\begin{smallmatrix}}
\newcommand{\esm}{\end{smallmatrix}}
\newcommand{\bspm}{\left(\begin{smallmatrix}}
\newcommand{\espm}{\end{smallmatrix}\right)}
\newcommand{\beq}{\begin{equation}}
\newcommand{\eeq}{\end{equation}}
\def\bar{\overline}

\def\hat{\widehat}

\def\^{\wedge}

\def\C{\mathbbm{C}}

\def\cH{{\mathcal H}}

\def\cN{{\mathcal N}}

\def\cT{{\mathcal T}}

\def\Z{\mathbbm{Z}}

\def\G{{\Gamma}}

\def\m{{\mu}}

\usepackage{tikz}
\usetikzlibrary{arrows,snakes,shapes.arrows,decorations.markings}
     \tikzset{>=triangle 90}
     \tikzstyle{bbc}=[draw,circle,fill=black,scale=.75]
     \tikzstyle{rc}=[circle,fill=red,scale=.6]
     \tikzstyle{wc}=[draw,circle,scale=.75]

\def\bar{\overline}

\def\hat{\widehat}

\def\^{\wedge}

\def\G{{\Gamma}}

\def\m{{\mu}}

\def\cH{{\mathcal H}}

\def\cN{{\mathcal N}}

\def\cT{{\mathcal T}}

\def\C{\mathbb{C}}

\def\Z{\mathbb{Z}} 

\def\beq{\begin{equation}}
\def\eeq{\end{equation}}

\newcommand{\bpmat}{\begin{pmatrix}}
\newcommand{\epmat}{\end{pmatrix}}
\newcommand{\bsmat}{\begin{smallmatrix}}
\newcommand{\esmat}{\end{smallmatrix}}
\newfloatcommand{capbtabbox}{table}[][\FBwidth]

\def\bz{\bar{z}}

\input{list_quivers}

\usepackage{todonotes}

\title{Magnetic quivers for rank 2 theories}
\author[1,2]{Antoine Bourget,}
\author[3]{ Julius F.\ Grimminger,}
\author[4,5]{ Mario Martone,}
\author[4,5,6]{ Gabi Zafrir}
\affiliation[1]{Université Paris-Saclay, CNRS, CEA, Institut de physique théorique, 91191, Gif-sur-Yvette, France}
\affiliation[2]{Laboratoire de Physique de l’\'Ecole normale supérieure, ENS, Université PSL, CNRS, Sorbonne
Université, Université de Paris, F-75005 Paris, France}
\affiliation[3]{Theoretical Physics Group, The Blackett Laboratory, Imperial College, London,
Prince Consort Road London, SW7 2AZ, UK}
\affiliation[4]{C.~N.~Yang Institute for Theoretical Physics,  Stony Brook University,Stony Brook, NY 11794-3840, USA}
\affiliation[5]{Simons Center for Geometry and Physics, Stony Brook University, Stony Brook, NY 11794-3840, USA}
\affiliation[6]{Dipartimento di Fisica, Università di Milano-Bicocca \& INFN, Sezione di Milano-Bicocca, I-20126 Milano, Italy}
\emailAdd{antoine.bourget@polytechnique.org}
\emailAdd{julius.grimminger17@imperial.ac.uk}
\emailAdd{mmartone@scgp.stonybrook.edu}
\emailAdd{gabi.zafrir@unimib.it}

\abstract{
In this note we construct magnetic quivers for the known rank-2 four dimensional $\cN=2$ superconformal field theories. For every rank-1 theory one can find a unitary magnetic quiver; we observe that this is no longer possible at rank 2. Our list of magnetic quivers necessarily includes orthosymplectic quivers, in addition to unitary ones, of both the simply and non-simply laced variety. Using quiver subtraction, one can compute Higgs branch Hasse diagrams and compare with the results obtained via other methods finding nearly perfect agreement. 
}

\begin{document}
\maketitle 

\section{Introduction and Summary}

The study of four dimensional supersymmetric theories has taught us a lot about the behavior of quantum field theory (QFT). Of particular interest in this study are theories with $\mathcal{N}=2$ supersymmetry (henceforth we will often drop the reference to the spacetime dimensionality, implicitly assuming that we consider four dimensional QFTs); on the one side, this set of theories is more constrained than the more phenomenologically viable $\mathcal{N}=1$ theories or theories with no supersymmetry altogether, and many exact results can be extracted about them. On the other side, these are not as severely constrained as the maximally supersymmetric, \emph{i.e.} $\mathcal{N}=4$, theories and the recently discovered $\mathcal{N}=3$ theories, thus exhibiting much richer behavior.

An important part in the study of $\mathcal{N}=2$ theories is the study of their moduli space of vacua. This can be split into two components or branches: the Coulomb and Higgs branch. For $\mathcal{N}=2$ theories with conformal symmetry, i.e. $\mathcal{N}=2$ superconformal field theories (SCFTs), which will be the focus of the present work, the $\mathrm{SU}(2)_R\times \mathrm{U}(1)_r$ R-symmetry differentiates the two. Specifically, the Coulomb branch (CB) and the Higgs branch (HB) can be identified as the part of the moduli space acted upon by the $\mathrm{U}(1)_r$ and $\mathrm{SU}(2)_R$ part respectively. In other words CB and HB operators can be characterized solely in terms of their properties in superconformal representation theory.

Representation theory is by itself very constraining but doesn't capture many properties which follow instead from a more geometric analysis. In the case of the CB, for example, the operators parametrizing it are in certain short representations of the $\mathcal{N}=2$ superconformal group\footnote{This type of multiplets is denoted as $\overline{\mathcal{E}}_{r(j,0)}$ by \cite{Dolan:2002zh} and as $L \overline{B}_1 [j;0]^{(0;r)}$ by \cite{Cordova:2016emh}.} for which the superconformal primary is an $\mathrm{SU}(2)_R$ singlet, has its angular momentum restricted, but not necessarily zero, and its dimension proportional to its $\mathrm{U}(1)_r$ charge, but not necessarily rational! Furthermore, there are no first principle constraints on its flavor representation. However, the picture is more constrained in actual physical theories; the primaries of CB operators must have trivial angular momentum \cite{Manenti:2019jds}, trivial flavor symmetry charges \cite{Buican:2014qla} and their dimension is restricted to be a rational number out of a list depending on the dimension of the CB \cite{Caorsi:2018zsq,Argyres:2018urp}.

These extra geometric constraints have been exploited for instance in attempts to classify possible CB geometries at low rank, notably for rank one in \cite{Argyres:2015ffa,Argyres:2015gha,Argyres:2016xua,Argyres:2016xmc} and generalizations with an eye on higher rank cases in \cite{Martone:2020nsy,Argyres:2020wmq}. The hope is to use these methods to classify 4d $\mathcal{N}=2$ SCFTs. 

Similarly, the operators associated with the HB are also in short representations of the $\mathcal{N}=2$ superconformal group\footnote{This type of multiplets is denoted as $\hat{\mathcal{B}}_R$ by \cite{Dolan:2002zh} and as $B_1 \overline{B}_1 [0;0]^{(R;0)}$ by \cite{Cordova:2016emh}.}. In this case superconformal symmetry restricts its superconformal primaries to be a $\mathrm{U}(1)_r$ singlet, to have trivial angular momentum and to have its dimension proportional to its $\mathrm{SU}(2)_R$ representation - which also forces the conformal dimension to be integer. HB operators are also known to carry global symmetry charges.

As it happens for the CB, there appear to exist extra geometric constraints restricting HB operators as well. That is there are many (hyper-K\"ahler) spaces which appear to be perfectly consistent with the aforementioned constraints, but are known to not correspond to any physically realized theory. Consider the $1$-instanton moduli space of a simple compact Lie group $G$. This is a hyper-K\"ahler space for any $G$, and there are known examples of $\mathcal{N}=2$ SCFTs whose HB is such a space for specific groups $G$\footnote{For instance the Lagrangian SCFT based on gauge group $\mathrm{SU}(2)$ with four doublet hypermultiplets for $G=\mathrm{SO}(8)$}. However, if one  assumes that on a generic point of such a HB the low-energy theory is only the collection (of an appropriate number to account for the dimension of the HB) of free hypers, one can determine the anomalies of the $4d$ SCFT from those of said free hypers. This only works if the group $G$ is one of a handful of cases \cite{Shimizu:2017kzs}. Similar results have also been found using bootstrap methods, see \cite{Beem:2013sza,Lemos:2015orc}. Furthermore, the case of $G=G_2$ or $F_4$ are allowed by the previously mentioned reasoning however no $\mathcal{N}=2$ SCFT with that moduli space has been found.

These issues motivate a more careful study of the question of what (hyper-K\"ahler) spaces can be realized as the HBs of $\mathcal{N}=2$ SCFTs. A standard approach to characterize $\mathcal{N}=2$ HBs is the well known \emph{hyper-K\"ahler quotient}. But these techniques assume a Lagrangian description which isn't available in the majority of cases. Another, more recent, method is to realize HBs using the CB of $3d$ $\mathcal{N}=4$ quiver theories, which are  referred to as \emph{magnetic quivers}. This later approach appears to be particularly suited for our purposes, as many of the so-called non-Lagrangian $\mathcal{N}=2$ SCFTs have Lagrangian 3d mirrors when reduced to 3d. We are then led to reformulate our initial question as identifying what magnetic quiver can describe the HB of $\mathcal{N}=2$ SCFTs; Can magnetic quivers describe the HBs of all $\mathcal{N}=2$ SCFTs? If so what are the necessary building blocks of the magnetic quivers that are required to achieve this? 

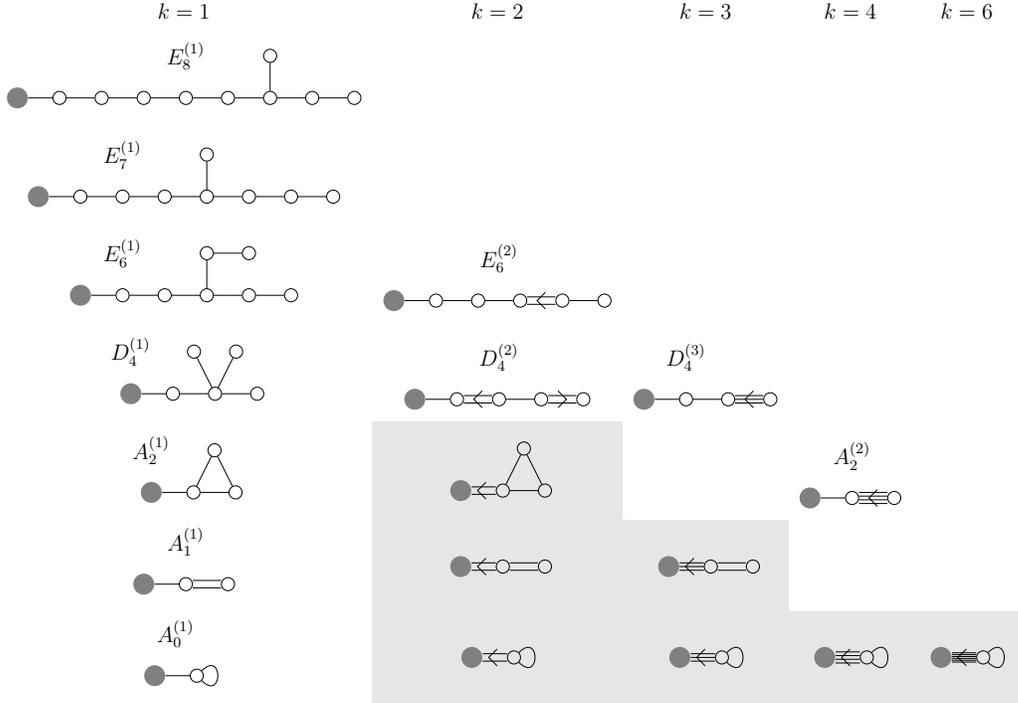
\begin{figure}
\begin{center}
     \hspace*{-1cm}\scalebox{.7}{\begin{tabular}{ccccc}  
   $k=1$ &   $k=2$ &   $k=3$ &   $k=4$ &   $k=6$    \\
\raisebox{-.5\height}{\begin{tikzpicture}[x=.8cm,y=.8cm,decoration={markings,mark=at position 0.5 with {\arrow{>},arrowhead=1cm}}]
\node at (1,1.5) {};
\node at (1,-.5) {};
\node at (5,1) {$E_8^{(1)}$};
\node (g1) at (1,0) [biggauge] {};
\node (g2) at (2,0) [gauge] {};
\node (g3) at (3,0) [gauge] {};
\node (g4) at (4,0) [gauge] {};
\node (g5) at (5,0) [gauge] {};
\node (g6) at (6,0) [gauge] {};
\node (g7) at (7,0) [gauge] {};
\node (g8) at (8,0) [gauge] {};
\node (g9) at (9,0) [gauge] {};
\node (g10) at (7,1) [gauge] {};
\draw (g1)--(g2)--(g3)--(g4)--(g5)--(g6)--(g7)--(g8)--(g9) (g7)--(g10);
\end{tikzpicture}} & &  & & \\   
\raisebox{-.5\height}{\begin{tikzpicture}[x=.8cm,y=.8cm,decoration={markings,mark=at position 0.5 with {\arrow{>},arrowhead=1cm}}]
\node at (4,1.5) {};
\node at (4,-.5) {};
\node at (4,1) {$E_7^{(1)}$};
\node (g2) at (2,0) [biggauge] {};
\node (g3) at (3,0) [gauge] {};
\node (g4) at (4,0) [gauge] {};
\node (g5) at (5,0) [gauge] {};
\node (g6) at (6,0) [gauge] {};
\node (g7) at (7,0) [gauge] {};
\node (g8) at (8,0) [gauge] {};
\node (g9) at (9,0) [gauge] {};
\node (g10) at (6,1) [gauge] {};
\draw (g2)--(g3)--(g4)--(g5)--(g6)--(g7)--(g8)--(g9) (g6)--(g10);
\end{tikzpicture}} & &  & & \\  
 \raisebox{-.5\height}{\begin{tikzpicture}[x=.8cm,y=.8cm,decoration={markings,mark=at position 0.5 with {\arrow{>},arrowhead=1cm}}]
\node at (4,1.5) {};
\node at (4,-.5) {};
\node at (4,1) {$E_6^{(1)}$};
\node (g3) at (3,0) [biggauge] {};
\node (g4) at (4,0) [gauge] {};
\node (g5) at (5,0) [gauge] {};
\node (g6) at (6,0) [gauge] {};
\node (g7) at (7,0) [gauge] {};
\node (g8) at (8,0) [gauge] {};
\node (g10) at (6,1) [gauge] {};
\node (g11) at (7,1) [gauge] {};
\draw (g3)--(g4)--(g5)--(g6)--(g7)--(g8) (g6)--(g10)--(g11);
\end{tikzpicture}}
& 
\raisebox{-.5\height}{\begin{tikzpicture}[x=.8cm,y=.8cm,decoration={markings,mark=at position 0.5 with {\arrow{>},arrowhead=1cm}}]
\node at (2.5,1) {$E_6^{(2)}$};
\node (g0) at (0,0) [biggauge] {};
\node (g1) at (1,0) [gauge] {};
\node (g2) at (2,0) [gauge] {};
\node (g3) at (3,0) [gauge] {};
\node (g4) at (4,0) [gauge] {};
\node (g5) at (5,0) [gauge] {};
\draw (g0)--(g1)--(g2)--(g3) (g4)--(g5);
\draw[transform canvas={yshift=2pt}] (g3)--(g4);
\draw[transform canvas={yshift=-2pt}] (g4)--(g3);
\draw (3.6,.2)--(3.4,0)--(3.6,-.2);
\end{tikzpicture}}
 &  & & \\ 
 \raisebox{-.5\height}{\begin{tikzpicture}[x=.8cm,y=.8cm,decoration={markings,mark=at position 0.5 with {\arrow{>},arrowhead=1cm}}]
\node at (4,1.5) {};
\node at (4,-.5) {};
\node at (3,1) {$D_4^{(1)}$};
\node (g3) at (3,0) [biggauge] {};
\node (g4) at (4,0) [gauge] {};
\node (g5) at (5,0) [gauge] {};
\node (g6) at (6,0) [gauge] {};
\node (g10) at (4.5,1) [gauge] {};
\node (g11) at (5.5,1) [gauge] {};
\draw (g3)--(g4)--(g5)--(g6) (g5)--(g10) (g5)--(g11);
\end{tikzpicture}} & 
\raisebox{-.5\height}{\begin{tikzpicture}[x=.8cm,y=.8cm,decoration={markings,mark=at position 0.5 with {\arrow{>},arrowhead=1cm}}]
\node (g2) at (2,0) [biggauge] {};
\node at (4,1) {$D_4^{(2)}$};
\node (g3) at (3,0) [gauge] {};
\node (g4) at (4,0) [gauge] {};
\node (g5) at (5,0) [gauge] {};
\node (g6) at (6,0) [gauge] {};
\draw (g2)--(g3) (g4)--(g5);
\draw[transform canvas={yshift=2pt}] (g3)--(g4);
\draw[transform canvas={yshift=-2pt}] (g4)--(g3);
\draw (3.6,.2)--(3.4,0)--(3.6,-.2);
\draw[transform canvas={yshift=2pt}] (g5)--(g6);
\draw[transform canvas={yshift=-2pt}] (g5)--(g6);
\draw (5.4,.2)--(5.6,0)--(5.4,-.2);
\end{tikzpicture}} & 
\raisebox{-.5\height}{\begin{tikzpicture}[x=.8cm,y=.8cm,decoration={markings,mark=at position 0.5 with {\arrow{>},arrowhead=1cm}}]
\node at (2,1) {$D_4^{(3)}$};
\node (g1) at (1,0) [biggauge] {};
\node (g2) at (2,0) [gauge] {};
\node (g3) at (3,0) [gauge] {};
\node (g4) at (4,0) [gauge] {};
\draw (g1)--(g2)--(g3);
\draw[transform canvas={yshift=2pt}] (g3)--(g4);
\draw[transform canvas={yshift=0pt}] (g4)--(g3);
\draw[transform canvas={yshift=-2pt}] (g4)--(g3);
\draw (3.6,.2)--(3.4,0)--(3.6,-.2);
\end{tikzpicture}}   & & \\  
 \raisebox{-.5\height}{\begin{tikzpicture}[x=.8cm,y=.8cm,decoration={markings,mark=at position 0.5 with {\arrow{>},arrowhead=1cm}}]
\node at (3,1) {$A_2^{(1)}$};
\node at (4,1.5) {};
\node at (4,-.5) {};
\node (g3) at (3,0) [biggauge] {};
\node (g4) at (4,0) [gauge] {};
\node (g5) at (5,0) [gauge] {};
\node (g10) at (4.5,1) [gauge] {};
\draw (g3)--(g4)--(g5)--(g10)--(g4);
\end{tikzpicture}} & \cellcolor{black!10}
\raisebox{-.5\height}{\begin{tikzpicture}[x=.8cm,y=.8cm,decoration={markings,mark=at position 0.5 with {\arrow{>},arrowhead=1cm}}]
\node (g3) at (3,0) [biggauge] {};
\node (g4) at (4,0) [gauge] {};
\node (g5) at (5,0) [gauge] {};
\node (g6) at (4.5,1) [gauge] {};
\draw (g4)--(g5)--(g6)--(g4);
\draw[transform canvas={yshift=2pt}] (g3)--(g4);
\draw[transform canvas={yshift=-2pt}] (g4)--(g3);
\draw (3.6,.2)--(3.4,0)--(3.6,-.2);
\end{tikzpicture}} &  &   \raisebox{-.5\height}{\begin{tikzpicture}[x=.8cm,y=.8cm,decoration={markings,mark=at position 0.5 with {\arrow{>},arrowhead=1cm}}]
\node at (3,1) {$A_2^{(2)}$};
\node (g2) at (2,0) [biggauge] {};
\node (g3) at (3,0) [gauge] {};
\node (g4) at (4,0) [gauge] {};
\draw (g2)--(g3);
\draw[transform canvas={yshift=3pt}] (g4)--(g3);
\draw[transform canvas={yshift=1pt}] (g4)--(g3);
\draw[transform canvas={yshift=-1pt}] (g4)--(g3);
\draw[transform canvas={yshift=-3pt}] (g4)--(g3);
\draw (3.6,.2)--(3.4,0)--(3.6,-.2);
\end{tikzpicture}} & \\ 
\raisebox{-.5\height}{\begin{tikzpicture}[x=.8cm,y=.8cm,decoration={markings,mark=at position 0.5 with {\arrow{>},arrowhead=1cm}}]
\node at (4,0.5) {};
\node at (4,1) {$A_1^{(1)}$};
\node at (4,-.5) {};
\node (g3) at (3,0) [biggauge] {};
\node (g4) at (4,0) [gauge] {};
\node (g5) at (5,0) [gauge] {};
\draw (g3)--(g4);
\draw[transform canvas={yshift=2pt}] (g4)--(g5);
\draw[transform canvas={yshift=-2pt}] (g4)--(g5);
\end{tikzpicture}}
 & \cellcolor{black!10}
\raisebox{-.5\height}{\begin{tikzpicture}[x=.8cm,y=.8cm,decoration={markings,mark=at position 0.5 with {\arrow{>},arrowhead=1cm}}]
\node (g3) at (3,0) [biggauge] {};
\node (g4) at (4,0) [gauge] {};
\node (g5) at (5,0) [gauge] {};
\draw[transform canvas={yshift=2pt}] (g3)--(g4);
\draw[transform canvas={yshift=-2pt}] (g4)--(g3);
\draw (3.6,.2)--(3.4,0)--(3.6,-.2);
\draw[transform canvas={yshift=2pt}] (g5)--(g4);
\draw[transform canvas={yshift=-2pt}] (g4)--(g5);
\end{tikzpicture}} &  \cellcolor{black!10}
\raisebox{-.5\height}{\begin{tikzpicture}[x=.8cm,y=.8cm,decoration={markings,mark=at position 0.5 with {\arrow{>},arrowhead=1cm}}]
\node (g3) at (3,0) [biggauge] {};
\node (g4) at (4,0) [gauge] {};
\node (g5) at (5,0) [gauge] {};
\draw[transform canvas={yshift=2pt}] (g3)--(g4);
\draw[transform canvas={yshift=0pt}] (g4)--(g3);
\draw[transform canvas={yshift=-2pt}] (g4)--(g3);
\draw (3.6,.2)--(3.4,0)--(3.6,-.2);
\draw[transform canvas={yshift=2pt}] (g5)--(g4);
\draw[transform canvas={yshift=-2pt}] (g4)--(g5);
\end{tikzpicture}}  & & \\  
\raisebox{-.5\height}{\begin{tikzpicture}[x=.8cm,y=.8cm,decoration={markings,mark=at position 0.5 with {\arrow{>},arrowhead=1cm}}]
\node at (3.5,1) {$A_0^{(1)}$};
\node (g3) at (3,0) [biggauge] {};
\node (g4) at (4,0) [gauge] {};
\draw (g3)--(g4);
\draw (g4) to [out=45,in=315,looseness=8] (g4);
\end{tikzpicture}} & \cellcolor{black!10} \raisebox{-.5\height}{\begin{tikzpicture}[x=.8cm,y=.8cm,decoration={markings,mark=at position 0.5 with {\arrow{>},arrowhead=1cm}}]
\node (g3) at (3,0) [biggauge] {};
\node (g4) at (4,0) [gauge] {};
\draw[transform canvas={yshift=2pt}] (g3)--(g4);
\draw[transform canvas={yshift=-2pt}] (g4)--(g3);
\draw (3.6,.2)--(3.4,0)--(3.6,-.2);
\draw (g4) to [out=45,in=315,looseness=8] (g4);
\end{tikzpicture}} & \cellcolor{black!10} \raisebox{-.5\height}{\begin{tikzpicture}[x=.8cm,y=.8cm,decoration={markings,mark=at position 0.5 with {\arrow{>},arrowhead=1cm}}]
\node (g3) at (3,0) [biggauge] {};
\node (g4) at (4,0) [gauge] {};
\draw[transform canvas={yshift=2pt}] (g3)--(g4);
\draw[transform canvas={yshift=0pt}] (g4)--(g3);
\draw[transform canvas={yshift=-2pt}] (g4)--(g3);
\draw (3.6,.2)--(3.4,0)--(3.6,-.2);
\draw (g4) to [out=45,in=315,looseness=8] (g4);
\end{tikzpicture}} & \cellcolor{black!10} \raisebox{-.5\height}{\begin{tikzpicture}[x=.8cm,y=.8cm,decoration={markings,mark=at position 0.5 with {\arrow{>},arrowhead=1cm}}]
\node (g3) at (3,0) [biggauge] {};
\node (g4) at (4,0) [gauge] {};
\draw[transform canvas={yshift=3pt}] (g3)--(g4);
\draw[transform canvas={yshift=1pt}] (g4)--(g3);
\draw[transform canvas={yshift=-1pt}] (g4)--(g3);
\draw[transform canvas={yshift=-3pt}] (g4)--(g3);
\draw (3.6,.2)--(3.4,0)--(3.6,-.2);
\draw (g4) to [out=45,in=315,looseness=8] (g4);
\end{tikzpicture}} & \cellcolor{black!10} \raisebox{-.5\height}{\begin{tikzpicture}[x=.8cm,y=.8cm,decoration={markings,mark=at position 0.5 with {\arrow{>},arrowhead=1cm}}]
\node (g3) at (3,0) [biggauge] {};
\node (g4) at (4,0) [gauge] {};
\draw[transform canvas={yshift=2.5pt}] (g3)--(g4);
\draw[transform canvas={yshift=1.5pt}] (g3)--(g4);
\draw[transform canvas={yshift=0.5pt}] (g3)--(g4);
\draw[transform canvas={yshift=-.5pt}] (g3)--(g4);
\draw[transform canvas={yshift=-1.5pt}] (g3)--(g4);
\draw[transform canvas={yshift=-2.5pt}] (g3)--(g4);
\draw (3.6,.2)--(3.4,0)--(3.6,-.2);
\draw (g4) to [out=45,in=315,looseness=8] (g4);
\end{tikzpicture}} \\ 
\end{tabular}}
\end{center}
    \caption{Patterns that appear in some magnetic quivers of 4d $\mathcal{N}=2$ SCFTs, organized in columns labeled by the folding degree $k$. In all cases, the gray dot represents the rest of the quiver. When labeled (white cells), the white dots form the corresponding (twisted) affine Dynkin diagram. The first column corresponds to the simply laced algebras in the Deligne-Cvitanovi\'c sequence. In the gray cells, the Dynkin diagrams are not folded, but the whole diagram becomes long compared to the rest of the quiver. }
    \label{fig:patterns}
\end{figure}

A first step towards answering these difficult general questions is to look at the HBs of known $\mathcal{N}=2$ SCFTs and identifying the corresponding magnetic quivers. To do this systematically, it is convenient to organize the $\mathcal{N}=2$ SCFTs based on their rank, i.e. the complex dimension of their CBs. The study of the simplest case - rank 1 - was already carried out in \cite{Bourget:2020asf}. It was found that all Higgs branches of rank 1 theories can be described by a magnetic quiver with unitary gauge nodes, possibly non simply laced, with shapes following the patterns of Figure \ref{fig:patterns}. 

Here we perform the study of the next simplest case, namely rank-2 $\mathcal{N}=2$ SCFTs. A list of 69 known rank 2 $\mathcal{N}=2$ SCFTs has been compiled in \cite{Martone:2021ixp}; in the following, we attempt to provide magnetic quivers for every theory in that list (we follow the numbering of these 69 theories as given in \cite[Tables 1-3]{Martone:2021ixp}). The results are presented in Tables \ref{tab:listMagneticQuivers1} to \ref{tab:listMagneticQuivers6}. We summarize here the salient features of our analysis:
\begin{itemize}
    \item A majority of theories (about 88 \%, i.e. 61 out of the 69 theories) admit a unitary magnetic quiver, among which 82 \% (50 out of 61) follow the patterns of Figure \ref{fig:patterns}. The unitary quivers are simply laced (34 \%) or non simply laced (66 \%). 
    \item For four theories, it appears necessary to resort to orthosymplectic quivers (simply or non simply laced). 
    \item For four theories, we are not able to provide a magnetic quiver. This is due, in some cases, to the lack of brane systems describing these theories, and in others, when the brane system does exist, to a lack of techniques allowing to extract a magnetic quiver. This might be seen as an indication that one has to go beyond unitary-orthosymplectic quivers with fundamental and rank-2 representations. 
\end{itemize}
The quivers reflect the effects of RG flows triggered by mass deformations of the SCFTs \cite{Bourget:2020mez,vanBeest:2021xyt}. For the rank 2 theories, these flows have been charted in \cite{Martone:2021drm}, and we depict them with magnetic quivers in Figures \ref{fig:e8so20} to \ref{fig:su2}. 
When a theory admits a unitary magnetic quiver, one can use quiver subtraction to compute the Hasse diagram of symplectic leaves of the Higgs branch. The results of these computations are shown in Figures \ref{fig:HasseDiagramsTable1} to \ref{fig:HasseDiagramsTable4}. It should be noted that the class of quivers considered here pushes the quiver subtraction algorithms to the current limits of our understanding, and in some cases one has to resort to physical intuition to disambiguate the Hasse diagram. It would be highly desirable to cross check the results obtained here using other methods.

In the rest of this note we describe the methods used to derive the magnetic quivers, and the Hasse diagrams, and we perform a variety of consistency checks, namely: ($i$) agreement with the essential $\cN=2$ data, \emph{i.e.} dimension of the HB and flavor symmetry ($ii$) agreement with detailed HB stratification independently derived ($iii$) agreement, when available, with HB Hilbert series calculation. We comment in particular on the cases where no unitary magnetic quiver was found. Appendix \ref{appendix5dWebs} gives details on the 5d brane webs which are the main tool used to derive the quivers, and Appendix \ref{appendixHasse} contains the Hasse diagrams derived from said quivers by quiver subtraction. Appendices \ref{appendixHLindex} and \ref{app:N=3} focus on the interesting cases of the $\mathfrak{sp}(4)_7 \times \mathfrak{sp}(8)_8$ theory and $\mathcal{N}=3$ theories, respectively.

Before delving into the details, let us finally mention that it is remarkable that many quivers can be guessed, before any computation is made, based purely on data from the SCFT. This can be seen as one more indication of the many restrictions that constrain 4d $\mathcal{N}=2$ SCFTs, and of their geometric `simplicity', which is apparent in the corresponding `simplicity' of most of the magnetic quivers. It is however tempting to expect that this simplicity is a consequence of the low rank of the SCFTs, and that at higher ranks a larger and larger proportion of Higgs branch geometries won't admit such an elementary description.

\section{Magnetic quivers of rank-2}
\label{sec:quivers}

In this section, we give magnetic quivers for most of the rank-2 4d $\mathcal{N}=2$ SCFTs identified in \cite{Martone:2021ixp}. We recall that for a given theory $\mathcal{T}$, a quiver $\mathsf{Q}$ is said to be a \emph{magnetic quiver} for $\mathcal{T}$ if 
\begin{equation}
   \mathcal{H} \left( \mathcal{T} \right) = \mathcal{C}\left( \mathsf{Q} \right) \, , 
    \label{definitionMagneticQuiver}
\end{equation}
where $\mathcal{H} \left( \mathcal{T} \right)$ is the HB of the 4d theory $\mathcal{T}$ and $\mathcal{C}\left(\mathsf{Q}\right)$ denotes the 3d $\mathcal{N} = 4$ CB of the quiver $\mathsf{Q}$.\footnote{The quiver $\mathsf{Q}$ may be non-simply laced \cite{Cremonesi:2014xha,Nakajima:2019olw,Bourget:2021xex}, in this case it does not define a gauge theory, however the CB is well defined.}
The definition (\ref{definitionMagneticQuiver}) does not imply that a magnetic quiver for a given theory is unique. To take a concrete example, the rank-1 $E_6$ Minahan-Nemeschanski theory has the closure of the minimal nilpotent orbit of $E_6$ as its HB, for which many magnetic quivers are known: the affine $E_6$ diagram, a twisted affine version \cite[Table 12]{Bourget:2020asf}, two orthosymplectic quivers \cite[Table 1]{Bourget:2020gzi} and \cite[Eq (6.3)]{Bourget:2020xdz}, and a folded orthosymplectic quiver \cite[Table 2]{Bourget:2021xex}.

It should be noted that the classical HB of a $4d$ $\mathcal{N}=2$ theory can be a non-reduced scheme\footnote{This can happen e.g.\ for SQCD, discussed at length in \cite{Bourget:2019rtl}.}, i.e.\ there is an operator $A\neq0$ in the classical HB chiral ring, s.t.\ $A^n=0$, $n\in \mathbb{N}$. In this case the magnetic quiver only provides the reduced part. It is conjectured in \cite[Conjecture 1]{Beem:2017ooy}, that for an SCFT the HB is always reduced. If this is true, then the magnetic quiver gives a good characterization of the HB. A further complication is that the HB of a $4d$ $\mathcal{N}=2$ theory may consist of several cones which intersect non-trivially. In this case several magnetic quivers are needed to describe the HB, one for each cone. However, in all examples we know of in which the HB is not connected, the various cones in the classical HB are separated along the (quantum) CB, and the Higgs directions at any point of the CB form a single cone. The more singular the point in the CB the more Higgs directions are available. Because of scale invariance, this behavior is not allowed for SCFTs and we thus expect the HB to be a single cone. Indeed we know of no counterexample in $4d$.\footnote{Note that HBs of $5d$ $\mathcal{N}=1$ SCFTs may be non-reduced and consist of several cones \cite{Cremonesi:2015lsa}.} Note that the existence of several cones would imply the existence of two operators $A , B\neq 0$ in the HB chiral ring satisfying $AB=0$. 
When the HB of a 4d $\mathcal{N}=2$ SCFT is a single and reduced cone, the HB chiral ring is, almost by definition, an \emph{integral domain}.

Because of the non-unicity of a magnetic quiver, we make the following choices: 
\begin{enumerate}
    \item We provide a simply laced unitary quiver if we know one; 
    \item If not, we provide a non-simply laced unitary quiver if we know one; 
    \item If not, we provide an orthosymplectic quiver (which may be simply laced or not) if we know one; 
    \item If not, we do not give a magnetic quiver. 
\end{enumerate}
In most of this section, we focus on theories with precisely $\mathcal{N}=2$ supersymmetry, as when $\mathcal{N}>2$ the definition of a HB is somewhat artificial, it being part of a larger moduli space and we further comment on this issue in Section \ref{sec:N3}.
There are only four $\mathcal{N}=2$ theories in the list of \cite{Martone:2021ixp} for which we don't give a magnetic quiver, and we comment on them later in this section. 

In order to compute the magnetic quivers, we can use a host of different methods. For some theories, several methods are available; in those cases, every method gives the same result. The results are presented in Tables \ref{tab:listMagneticQuivers1}-\ref{tab:listMagneticQuivers6}. The last column indicates which of the following methods can be used to find the given magnetic quiver. 
\begin{enumerate}
    \item \label{webs} \textbf{Compactification from 5d}. By far the most powerful method is to use the realization of the 4d theories as (possibly twisted) compactification of 5d theories \cite{Martone:2021drm}. These 5d theories can often be realized as the world-volume of type IIB 5-brane webs, from which magnetic quivers can be deduced using web decomposition and intersection in tropical geometry as first explained in \cite{Cabrera:2018jxt} and later generalized in \cite[Appendix B]{Bourget:2020mez} for brane webs without orientifolds, and discussed in \cite{Bourget:2020gzi} and \cite{Akhond:2020vhc} with O5 planes. Equivalently, one can use generalized toric polygons \cite{Benini:2009gi} to concisely encode these brane webs and compute magnetic quivers \cite{vanBeest:2020kou,vanBeest:2020civ}. This is the format we use in the present paper in appendix \ref{appendix5dWebs}, where all the details are provided. 
    \item \label{classS} \textbf{Class S theories of type A and D}. Several theories are realized as class S on a sphere with punctures. For type $A$ with regular untwisted punctures, the magnetic quiver is star-shaped unitary \cite{Benini:2010uu}. For type $A$ with twisted punctures \cite{Beratto:2020wmn} and for type $D$, the magnetic quiver is (unitary-)orthosymplectic. 
    We use this construction for theory \# 26 which uses twisted $A_3$ punctures \cite{Chacaltana:2012ch}, and for theory \# 41 which uses $\mathbb{Z}_2$ twisted $D_4$ punctures \cite{Chacaltana:2013oka}. We return to those theories in section \ref{subsectionNonUnitary}. 
    \item \label{instanton} \textbf{Instanton Moduli Spaces}. Certain theories have HBs that can be identified with 2-instanton moduli spaces on $\mathbb{C}^2$ \cite{Shimizu:2017kzs,Beem:2019snk}, and magnetic quivers for these are given in \cite{Cremonesi:2014xha}. 
    \item \label{Sfolds} \textbf{S-fold theories}. S-folds were initially introduced as a way to construct $\mathcal{N}=3$ theories in 4d \cite{Garcia-Etxebarria:2015wns,Aharony:2016kai} and later $\mathcal{N}=2$ theories \cite{Apruzzi:2020pmv,Giacomelli:2020jel,Heckman:2020svr,Giacomelli:2020gee}. Magnetic quivers for those theories were constructed in \cite{Bourget:2020mez}. 
    \item \label{AD} \textbf{Argyres-Douglas theories}. Magnetic quivers for Argyres-Douglas theories (and generalizations) have been investigated by many authors over the past few years using dimensional reduction to 3d and 3d mirror symmetry \cite{Song:2017oew,Dedushenko:2019mnd,Beratto:2020wmn,Closset:2020scj,Giacomelli:2020ryy,Xie:2021ewm}.  
\end{enumerate}

\subsection{RG flows}
Many of the rank-2 SCFTs are related by RG flows, as worked out in \cite[Fig. 1]{Martone:2021drm}. These RG flows correspond to deformations of the HBs, which are depicted using magnetic quivers in Figures \ref{fig:e8so20}-\ref{fig:su6}. It should be noted that the quivers in those Figures are not related by any fully known graph theoretic algorithm -- in particular, they are not related by the quiver subtraction algorithm used to determine the Hasse diagram of symplectic leaves in a given HB. However, the 3d theories encoded by the magnetic quivers should be related in general by turning on Fayet-Iliopoulos (FI) parameters, as the HBs along the flows are in general obtained by turning on masses for hypermultiplets in an effective Lagrangian description. This requires understanding the precise mapping between the mass parameters of the 4d theory and the FI parameters of the 3d quivers, which may be quite subtle at times due to the decoupling of an overall $\mathrm{U}(1)$ factor in the 3d quivers, see \cite{vanBeest:2021xyt} for a recent exploration of this issue. Once the mapping is understood, graph theoretic methods involving subtraction of finite Dynkin diagrams can be used to deduce magnetic quivers after RG flow, along the lines of \cite{Bourget:2020mez,vanBeest:2021xyt}. 

Without entering into those details, certain striking patterns can be noted, for example the tail of many MQs can be linked with the well-known Deligne-Cvitanovi\'c sequence, which appears in studies of 4d $\mathcal{N}=2$ SCFTs \cite{Beem:2013sza,arakawa2018joseph,Beem:2019tfp}.  The simply laced members of the sequence are
\begin{equation}
    E_8 \longrightarrow E_7 \longrightarrow E_6 \longrightarrow D_4 \longrightarrow A_2 \longrightarrow A_1   \, . 
\end{equation}
The corresponding patterns in magnetic quivers are shown in Figure \ref{fig:patterns}, along with their non simply laced counterparts.  All rank 1 theories can be described with these patterns but only about 70 \% of the rank 2 theories listed in this paper.  

\subsection{Orthosymplectic quivers}
\label{subsectionNonUnitary}

In this subsection, we use the patterns identified above to argue for the absence of \emph{unitary} magnetic quivers for certain theories. Consider for instance the $\mathfrak{su}(2)_{5} \times \mathfrak{sp}(6)_{6} \times \mathfrak{u}(1)$ theory. The rank of the global symmetry being 5, the number of unitary nodes is 5 or 6, depending on whether there is a long $\mathrm{U}(1)$ node or not. Using the above pattern and the location of that theory in the RG-flow chart of Figure \ref{fig:sp12sp8f4}, a hypothetical magnetic quiver should then have structure\footnote{Note that we have indicated in white the nodes which are balanced, in order to reproduce the expected flavor symmetry. In principle there could be symmetry enhancement (see \cite{Gledhill:2021cbe} for a recent survey) and the quiver could have fewer balanced nodes. We assume here that this doesn't happen.   } 
\begin{equation}
\raisebox{-.5\height}{\begin{tikzpicture}[x=.8cm,y=.8cm,decoration={markings,mark=at position 0.5 with {\arrow{>},arrowhead=1cm}}]
\node (g2) at (2,0) [gauge] {};
\node (g3) at (3,0) [gauge] {};
\node (g4) at (4,0) [gauge] {};
\node (g5) at (5,0) [gaugeb] {};
\node (g6) at (6,0) [gauge] {};
\draw (g2)--(g3) (g4)--(g5);
\draw[transform canvas={yshift=2pt}] (g3)--(g4);
\draw[transform canvas={yshift=-2pt}] (g4)--(g3);
\draw (3.6,.2)--(3.4,0)--(3.6,-.2);
\draw[transform canvas={yshift=2pt}] (g5)--(g6);
\draw[transform canvas={yshift=-2pt}] (g5)--(g6);
\draw (5.4,.2)--(5.6,0)--(5.4,-.2);
\end{tikzpicture}}
\qquad \textrm{or} \qquad 
\raisebox{-.5\height}{\begin{tikzpicture}[x=.8cm,y=.8cm,decoration={markings,mark=at position 0.5 with {\arrow{>},arrowhead=1cm}}]
\node (g1) at (1,0) [gaugeb] {};
\node (g2) at (2,0) [gauge] {};
\node (g3) at (3,0) [gauge] {};
\node (g4) at (4,0) [gauge] {};
\node (g5) at (5,0) [gaugeb] {};
\node (g6) at (6,0) [gauge] {};
\draw (g1)--(g2)--(g3) (g4)--(g5);
\draw[transform canvas={yshift=2pt}] (g3)--(g4);
\draw[transform canvas={yshift=-2pt}] (g4)--(g3);
\draw (3.6,.2)--(3.4,0)--(3.6,-.2);
\draw[transform canvas={yshift=2pt}] (g5)--(g6);
\draw[transform canvas={yshift=-2pt}] (g5)--(g6);
\draw (5.4,.2)--(5.6,0)--(5.4,-.2);
\end{tikzpicture}}
\label{candidatesQuivers}
\end{equation}
The black nodes denote overbalanced nodes. The quaternionic dimension of the HB is 11, so the sum of the labels of these quivers has to be 12. 

Consider the second quiver in (\ref{candidatesQuivers}). The balance conditions and the dimension condition give five equations for the five unknown ranks, thus giving a unique solution, which has non-integer ranks, as shown in the red quiver below: 
\begin{equation}
\begin{tabular}{r}
\raisebox{-.5\height}{\begin{tikzpicture}[x=.8cm,y=.8cm,decoration={markings,mark=at position 0.5 with {\arrow{>},arrowhead=1cm}}]
\node (g2) at (2,0) [gauge,label=below:{$1$}] {};
\node (g3) at (3,0) [gaugeb,label=below:{$2$}] {};
\node (g4) at (4,0) [gauge,label=below:{$4$}] {};
\node (g5) at (5,0) [gauge,label=below:{$4$}] {};
\node (g6) at (6,0) [gauge,label=below:{$2$}] {};
\draw (g2)--(g3) (g4)--(g5);
\draw[transform canvas={yshift=2pt}] (g3)--(g4);
\draw[transform canvas={yshift=-2pt}] (g4)--(g3);
\draw (3.6,.2)--(3.4,0)--(3.6,-.2);
\draw[transform canvas={yshift=2pt}] (g5)--(g6);
\draw[transform canvas={yshift=-2pt}] (g5)--(g6);
\draw (5.4,.2)--(5.6,0)--(5.4,-.2);
\end{tikzpicture}} \\ 
\color{red}{\raisebox{-.5\height}{\begin{tikzpicture}[x=.8cm,y=.8cm,decoration={markings,mark=at position 0.5 with {\arrow{>},arrowhead=1cm}}]
\node (g2) at (2,0) [gauge,label=below:{$\frac{4}{3}$}] {};
\node (g3) at (3,0) [gauge,label=below:{$\frac{8}{3}$}] {};
\node (g4) at (4,0) [gauge,label=below:{$4$}] {};
\node (g5) at (5,0) [gaugeb,label=below:{$\frac{8}{3}$}] {};
\node (g6) at (6,0) [gauge,label=below:{$\frac{4}{3}$}] {};
\draw (g2)--(g3) (g4)--(g5);
\draw[transform canvas={yshift=2pt}] (g3)--(g4);
\draw[transform canvas={yshift=-2pt}] (g4)--(g3);
\draw (3.6,.2)--(3.4,0)--(3.6,-.2);
\draw[transform canvas={yshift=2pt}] (g5)--(g6);
\draw[transform canvas={yshift=-2pt}] (g5)--(g6);
\draw (5.4,.2)--(5.6,0)--(5.4,-.2);
\end{tikzpicture}}} \\ 
\raisebox{-.5\height}{\begin{tikzpicture}[x=.8cm,y=.8cm,decoration={markings,mark=at position 0.5 with {\arrow{>},arrowhead=1cm}}]
\node (g1) at (1,0) [gauge,label=below:{$1$}] {};
\node (g2) at (2,0) [gauge,label=below:{$2$}] {};
\node (g3) at (3,0) [gauge,label=below:{$3$}] {};
\node (g4) at (4,0) [gauge,label=below:{$4$}] {};
\node (g5) at (5,0) [gaugeb,label=below:{$2$}] {};
\node (g6) at (6,0) [gauge,label=below:{$1$}] {};
\draw (g1)--(g2)--(g3) (g4)--(g5);
\draw[transform canvas={yshift=2pt}] (g3)--(g4);
\draw[transform canvas={yshift=-2pt}] (g4)--(g3);
\draw (3.6,.2)--(3.4,0)--(3.6,-.2);
\draw[transform canvas={yshift=2pt}] (g5)--(g6);
\draw[transform canvas={yshift=-2pt}] (g5)--(g6);
\draw (5.4,.2)--(5.6,0)--(5.4,-.2);
\end{tikzpicture}}
\end{tabular}
\end{equation}
Above and below that quiver, we reproduce the adjacent quivers in Figure \ref{fig:sp12sp8f4} which, recall, represents the quivers of the theories immediately adjacent to the theory we are analyzing, along RG-flow trajectories. This shows that the ranks of the putative magnetic quiver for the $\mathfrak{su}(2)_{5} \times \mathfrak{sp}(6)_{6} \times \mathfrak{u}(1)$ indeed interpolate between the ranks of neighbors in the RG-flow chart, but this interpolation would require a non-sensical quiver with fractional ranks. 

Investigation of the second candidate in (\ref{candidatesQuivers}) is not more fruitful. 
Calling the ranks of the nodes $n_1 , \dots , n_6$, the four balance conditions plus the dimension condition give five equations, allowing to express all the ranks in terms of $n_1$ as 
\begin{equation}
    n_2 =  \frac{5n_1}{9}+\frac{4}{3} \qquad n_3 = \frac{n_1}{9}+\frac{8}{3} \qquad n_4 = 4-\frac{n_1}{3} \qquad n_5 = \frac{8}{3}-\frac{8 n_1}{9} \qquad n_6 =  \frac{4}{3}-\frac{4 n_1}{9} \, . 
\end{equation}
In order for these quantities to be integers, one needs $n_1 = 3$ modulo $9$, and there is no such value that makes all the $n_i >0$. 
In view of these negative results, it is therefore natural to turn to a broader class of quivers, namely orthosymplectic quivers. Using the class S construction of the theory, a natural candidate is 
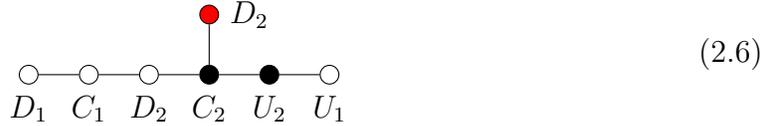
\begin{equation}
    \quiv{26} 
\end{equation}
In this quiver, the overbalanced nodes are denoted in black as usual, and an underbalanced node appears, denoted in red. Despite that node, it is possible to evaluate the CB Hilbert series for the 3d $\mathcal{N}=4$ theory using Hall-Littlewood techniques\footnote{We thank Rudolph Kalveks for the computation of the exact result.}, yielding 
\begin{equation}
   \frac{ \left( 
   \begin{array}{c}
      1 + 3 t + 23 t^2 + 92 t^3 + 410 t^4 + 1422 t^5 + 4828 t^6 +  \\
        14244 t^7 + 39757 t^8 + 100449 t^9 + 238641 t^{10} + 523542 t^{11} +  \\
        1081541 t^{12} + 2086065 t^{13} + 3799657 t^{14} + 6507468 t^{15} + \\
         10555585 t^{16} + 16175503 t^{17} + 23533981 t^{18} + 32452262 t^{19} + \\
         42567355 t^{20} + 53036689 t^{21} + 62940537 t^{22} + 71046632 t^{23} + \\
          76444660 t^{24} + 78287994 t^{25} + \textrm{palindrome} + t^{50}
   \end{array} \right)
  }{(1 - t)^{-3} (1 - t^2)^8 (1 - t^4)^8 (1 - t^3)^9}  = 1 + 25 t^2 +  \dots 
\end{equation}
As a check, the $t^2$ coefficient 25 matches with the expected dimension of the global symmetry, and the order of the pole at $t=1$ is 22, matching with the complex dimension of the HB. 

A similar analysis can be made for the $\mathfrak{sp}(4)_{7} \times \mathfrak{sp}(8)_{8} $ theory, showing that no unitary quiver following the pattern of Figure \ref{fig:patterns} can be found. We then also turn to orthosymplectic quivers, as shown in Table \ref{tab:listMagneticQuivers3}, but in this case, we actually need a folded orthosymplectic quiver \cite{Bourget:2021xex}. The argument goes as follows. We begin with the 5d realization of this theory proposed in \cite{Martone:2021drm}: the compactification of a certain 5d SCFT with a twist in a $\mathbb{Z}_2$ global symmetry. The specific 5d SCFT can be conveniently described as the UV completion of a 5d $\mathrm{SU}(4)$ gauge theory with an antisymmetric hypermultiplet, $8$ fundamental hypermultiplets and no Chern-Simons term. Here the $\mathbb{Z}_2$ symmetry we twist by acts on the gauge theory by charge conjugation. This 5d SCFT has a standard brane web description, see for instance \cite{Bergman:2015dpa}, so we would naively expect a unitary though non-simply laced quiver. However, as noted in \cite{Martone:2021drm}, this $\mathbb{Z}_2$ symmetry is not manifest in the brane construction, preventing the derivation of a magnetic quiver using this brane description.

Nevertheless, we can still exploit the 5d description to argue for a magnetic quiver thanks to the group theory coincidence that $\mathrm{SU}(4)=\mathrm{Spin}(6)$. This implies that we can describe the 5d SCFT by an alternative brane realization as a $\mathrm{Spin}(6)$ gauge theory with a vector hypermultiplet and $8$ spinor hypermultiplets, using the techniques in \cite{Zafrir:2015ftn}. Said brane realization then contains an O$5^-$ plane, and is of the form such that it reduces in 4d to a D type class S theory associated with a three punctured sphere, see \cite{Zafrir:2016jpu}. The web is given in the form of a generalized toric diagram, see (\ref{webTh23}). Interestingly, this 5d realization manifests the $\mathbb{Z}_2$ discrete symmetry, given by a reflection in the brane picture and the exchange of two identical punctures in the 4d class S picture. The latter gives a mirror orthosymplectic star-shaped quiver when reduced to 3d, with two identical legs. Similarly to the unitary case, we expect that folding the two identical legs will give a magnetic quiver describing the Higgs branch of the 4d theory we get by the $\mathbb{Z}_2$ twisted compactification of the associated 5d SCFT. This gives the non-simply laced orthosymplectic shown in Table \ref{tab:listMagneticQuivers3}. There are several indications that this magnetic quiver indeed describes the Higgs branch of the 4d $\mathfrak{sp}(4)_{7} \times \mathfrak{sp}(8)_{8} $ SCFT. First, the dimension of the Coulomb branch of the quiver matches the expected dimension of the Higgs branch of the 4d SCFT. Furthermore, one can exploit the 5d construction to motivate an expression for the Hall-Littlewood index of the 4d theory, which can then be compared against results obtained using the class S description. Finally, the Higgs branch dimension of the unfolded quiver turns out to be 3, in agreement with the expected dimension for that family. We refer the reader to appendix \ref{appendixHLindex} for the details.

\subsection{\texorpdfstring{A comment on $\mathcal{N} \geq 3$ theories}{A comment on N>=3 theories}}\label{sec:N3}

Two theories, in green in Table \ref{tab:listMagneticQuivers6}, have exactly $\mathcal{N}=3$ supersymmetry. They correspond to theories whose Coulomb branches are orbifolds defined via the complex reflection groups $G(3,1,2)$ and $G(4,1,2)$ -- such theories were classified in \cite{Argyres:2019ngz}. As reviewed in that article, one can use a Molien sum to compute the Higgs branch Hilbert series, since the Higgs branch can be realized as $\mathbb{C}^4 / (\Gamma \oplus \overline{\Gamma})$, where $\Gamma$ is an appropriate two dimensional representation of the complex reflection group. The resulting Hilbert series are: 
\begin{eqnarray}\label{HilbG312}
    G(3,1,2) \, : & \qquad & \frac{1-t+t^3+t^5-t^7+t^8}{(1-t
   ) \left(1-t^2\right)
   \left(1-t^3\right)
   \left(1-t^6\right)} \\ \label{HilbG412}
    G(4,1,2) \, : & \qquad & \frac{1-t^2+2
   t^4+2 t^8-t^{10}+t^{12}}{\left(1-t^2\right
   )^2 \left(1-t^4\right)
   \left(1-t^8\right)} \, . 
\end{eqnarray}
One can compare these results with the Hilbert series obtained from the monopole formula applied to the quivers presented in Table \ref{tab:listMagneticQuivers6}, finding perfect agreement. 
This constitutes a strong consistency check of the validity of the methods used throughout this paper. 

The same can be done for the four rank-2 $\mathcal{N}=4$ theories, outlined in blue in the tables. The gauge algebras are $A_2$, $B_2=C_2$, $D_2 = A_1 \oplus A_1$ and $G_2$. In those cases the reflection group is simply the corresponding Weyl group, and the Higgs branch Hilbert series evaluate to 
\begin{eqnarray}
    A_2 \, : & \qquad & \frac{1+t^2+2
   t^3+t^4+t^6}{\left(1-t^2\right)^2
   \left(1-t^3\right)^2} \\ 
    B_2 \, : & \qquad & \frac{1+t^2+4
   t^4+t^8+t^8}{\left(1-t^2\right)^2
   \left(1-t^4\right)^2} \\ 
    D_2 \, : & \qquad & \left( \frac{1+t^2}{\left(1-t^2
   \right)^2} \right)^2 \\ 
    G_2 \, : & \qquad & \frac{1 + t^2 + t^4+6
   t^6+t^8+t^{10}+t^{12}}{\left(1-t^2\right)^2
   \left(1-t^6\right)^2}  \, . 
\end{eqnarray}
One can compute the Coulomb branch Hilbert series for the magnetic quivers given in Tables \ref{tab:listMagneticQuivers3}, \ref{tab:listMagneticQuivers5}, \ref{tab:listMagneticQuivers6} and find agreement with the above results. Note that these quivers and their Hilbert series correspond to the $k_n$ slices defined in \cite[C.6]{Bourget:2020mez} as the 3d $\mathcal{N}=4$ Coulomb branch of the quivers 
\begin{equation}
    \begin{tikzpicture}
            \node[gauge,label=below:{$1$}] (1) at (0,0) {};
            \node[gauge,label=below:{$2$}] (2) at (1,0) {};
            \draw[very thick] (1)--(2);
            \draw (2) to [out=315,in=45,looseness=8] (2);
            \node at (0.5,0.3) {$n$};
            \draw (.6,.1)--(.4,0)--(.6,-.1);
    \end{tikzpicture} 
\end{equation}
with the correspondence 
\begin{equation}
    \begin{array}{c|cccc}
        \mathcal{N}=4 \textrm{ theory gauge algebra} &  D_2 & A_2 & B_2 & G_2 \\ \hline 
       \textrm{Corresponding } n  & n=2 &  n=3 &  n=4 &  n=6   
    \end{array}
\end{equation}

\subsection{Hasse diagrams}

The quiver subtraction algorithm explained in \cite{Cabrera:2018ann,Bourget:2019aer,Bourget:2020mez,Grimminger:2020dmg,Bourget:2020mez,Bourget:2021siw} can be used to compute a Hasse diagram from the magnetic quivers presented in Tables \ref{tab:listMagneticQuivers1}-\ref{tab:listMagneticQuivers6}. This gives information about the symplectic leaf structure and elementary slices of the HB, and therefore about the phase structure of the corresponding 4d SCFT. This can then be compared with the diagrams computed in \cite{Martone:2021ixp,Martone:2021drm}. The Hasse diagrams obtained from quiver subtraction are shown in Figures \ref{fig:HasseDiagramsTable1}-\ref{fig:HasseDiagramsTable4}. In most cases, the computation is straightforward, and there is direct agreement with the results of \cite{Martone:2021ixp,Martone:2021drm}. In this subsection we comment on some of the most interesting cases.

Our convention for slices are the following:
\begin{itemize}
    \item $a_n$, ... , $g_2$ are the minimal nilpotent orbit closures of the respective algebras.
    \item $h_{n,k}$ denotes the orbifolds $\mathbb{H}^n/\mathbb{Z}_k$ with charges $\pm1$ acting on the two $\mathbb{C}$s in each $\mathbb{H}$ factor.
    \item $\bar{h}_{n,k}$ is an elementary slice introduced in \cite[Sec. 3.3]{Bourget:2021siw}. Notable examples are $ac_n=\bar{h}_{n,2}$ and $ag_2=\bar{h}_{2,3}$, which appear in the affine Grassmannian of non-simply laced groups \cite{2003math......5095M}.
    \item $k_n$ denote the slices defined in \cite[C.6]{Bourget:2020mez}.
\end{itemize}

\begin{figure}
    \centering
    \begin{tikzpicture}
        \node (0) at (0,0) {$\begin{tikzpicture}
            \node[gauge,label=below:{$1$}] (1) at (1,0) {};
            \node[gauge,label=below:{$2$}] (2) at (2,0) {};
            \node[gauge,label=below:{$3$}] (3) at (3,0) {};
            \node[gauge,label=below:{$4$}] (4) at (4,0) {};
            \node[gauge,label=below:{$3$}] (5) at (5,0) {};
            \node[gauge,label=below:{$1$}] (6) at (6,0) {};
            \draw (1)--(2)--(3) (4)--(5);
		    \draw[transform canvas={yshift=1pt}] (3)--(4);
		    \draw[transform canvas={yshift=-1pt}] (3)--(4);
		    \draw (3.5+0.1,0.1)--(3.5-0.1,0)--(3.5+0.1,-0.1);
		    \draw[transform canvas={yshift=1pt}] (5)--(6);
		    \draw[transform canvas={yshift=-1pt}] (5)--(6);
		    \draw (5.5-0.1,0.1)--(5.5+0.1,0)--(5.5-0.1,-0.1);
        \end{tikzpicture}$
        };
        \node (l1) at (-3,-4.5) {$\begin{tikzpicture}
            \node[gauge,label=below:{$1$}] (5) at (5,0) {};
            \node[gauge,label=below:{$1$}] (6) at (6,0) {};
            \node[gauge,label=below:{$1$}] (7) at (7,0) {};
		    \draw[transform canvas={yshift=1pt}] (5)--(6);
		    \draw[transform canvas={yshift=-1pt}] (5)--(6);
		    \draw (5.5-0.1,0.1)--(5.5+0.1,0)--(5.5-0.1,-0.1);
		    \draw[transform canvas={yshift=4pt}] (6)--(7);
		    \draw[transform canvas={yshift=2pt}] (6)--(7);
		    \begin{scope}[yshift=3pt]
		    \draw (6.5+0.1,0.1)--(6.5-0.1,0)--(6.5+0.1,-0.1);
		    \end{scope}
		    \draw[transform canvas={yshift=-2pt}] (6)--(7);
		    \draw[transform canvas={yshift=-4pt}] (6)--(7);
		    \begin{scope}[yshift=-3pt]
		    \draw (6.5+0.1,0.1)--(6.5-0.1,0)--(6.5+0.1,-0.1);
		    \end{scope}
        \end{tikzpicture}$
        };
        \node (r1) at (3,-2.5) {$\begin{tikzpicture}
            \node[gauge,label=below:{$1$}] (1) at (1,0) {};
            \node[gauge,label=below:{$2$}] (2) at (2,0) {};
            \node[gauge,label=below:{$2$}] (3) at (3,0) {};
            \node[gauge,label=below:{$2$}] (4) at (4,0) {};
            \node[gauge,label=below:{$1$}] (5) at (5,0) {};
            \node[gauge,label=left:{$1$}] (6) at (2,1) {};
            \draw (1)--(2)--(3) (4)--(5);
		    \draw[transform canvas={yshift=1pt}] (3)--(4);
		    \draw[transform canvas={yshift=-1pt}] (3)--(4);
		    \draw (3.5+0.1,0.1)--(3.5-0.1,0)--(3.5+0.1,-0.1);
		    \draw[transform canvas={xshift=1pt}] (6)--(2);
		    \draw[transform canvas={xshift=-1pt}] (6)--(2);
		    \draw (2-0.1,0.5+0.1)--(2,0.5-0.1)--(2+0.1,0.5+0.1);
        \end{tikzpicture}$
        };
        \node (r2) at (3,-6.5) {$\begin{tikzpicture}
		    \node[gauge,label=left:{$1$}] (l) at (0,0) {};
		    \node[gauge,label=above:{$1$}] (u) at (2,1) {};
		    \node[gauge,label=below:{$1$}] (1d) at (1,-1) {};
		    \node[gauge,label=below:{$1$}] (d) at (2,-1) {};
		    \node[gauge,label=below:{$1$}] (2d) at (3,-1) {};
		    \node[gauge,label=right:{$1$}] (r) at (4,0) {};
		    \draw (l)--(u)--(r) (1d)--(d)--(2d);
		    \draw[transform canvas={xshift=1pt,yshift=1pt}] (l)--(1d);
		    \draw[transform canvas={xshift=-1pt,yshift=-1pt}] (l)--(1d);
		    \draw[transform canvas={xshift=1pt,yshift=-1pt}] (r)--(2d);
		    \draw[transform canvas={xshift=-1pt,yshift=1pt}] (r)--(2d);
		    \draw (0.5-0.3,-0.5)--(0.5,-0.5)--(0.5,-0.5+0.3);
		    \draw (3.5,-0.5+0.3)--(3.5,-0.5)--(3.5+0.3,-0.5);	
	    \end{tikzpicture}
	    $};
        \node (m) at (-1,-9) {$\begin{tikzpicture}
            \node[gauge,label=below:{$1$}] (3) at (3,0) {};
            \node[gauge,label=below:{$1$}] (4) at (4,0) {};
		    \draw[transform canvas={yshift=1pt}] (3)--(4);
		    \draw[transform canvas={yshift=-1pt}] (3)--(4);
        \end{tikzpicture}$
        };
        \node (b) at (-1,-11) {$\begin{tikzpicture}
            \node[gauge,label=below:{$1$}] (3) at (3,0) {};
        \end{tikzpicture}$
        };
        \draw[->] (0)--(l1);
        \draw[->] (l1)--(m);
        \draw[->] (0)--(r1);
        \draw[->] (r1)--(r2);
        \draw[->] (r2)--(m);
        \draw[->] (m)--(b);
        \node at (-2.5,-2.25) {$-e_6$};
        \node at (-2.5,-6.75) {$-A_1$};
        \node at (2,-0.75) {$-d_4$};
        \node at (3.5,-4.25) {$-c_3$};
        \node at (0.5,-8) {$-c_4$};
        \node at (-0.5,-10) {$-a_1$};
        \node at (9,-6) {$\begin{tikzpicture}
            \node[hasse] (0) at (0,0) {};
            \node[hasse] (l1) at (-1,-1.5) {};
            \node[hasse] (r1) at (1,-1) {};
            \node[hasse] (r2) at (1,-2) {};
            \node[hasse] (m) at (0,-3) {};
            \node[hasse] (b) at (0,-4) {};
            \draw (0)--(l1)--(m)--(b) (m)--(r2)--(r1)--(0);
            \node at (-0.8,-0.75) {$e_6$};
            \node at (-0.8,-2.25) {$A_1$};
            \node at (-0.3,-3.5) {$a_1$};
            \node at (0.8,-0.5) {$d_4$};
            \node at (1.3,-1.5) {$c_3$};
            \node at (0.8,-2.5) {$c_4$};
        \end{tikzpicture}$};
    \end{tikzpicture}
    \caption{Quiver subtraction for the magnetic quiver of theory \# 34}
    \label{fig:QS34}
\end{figure}
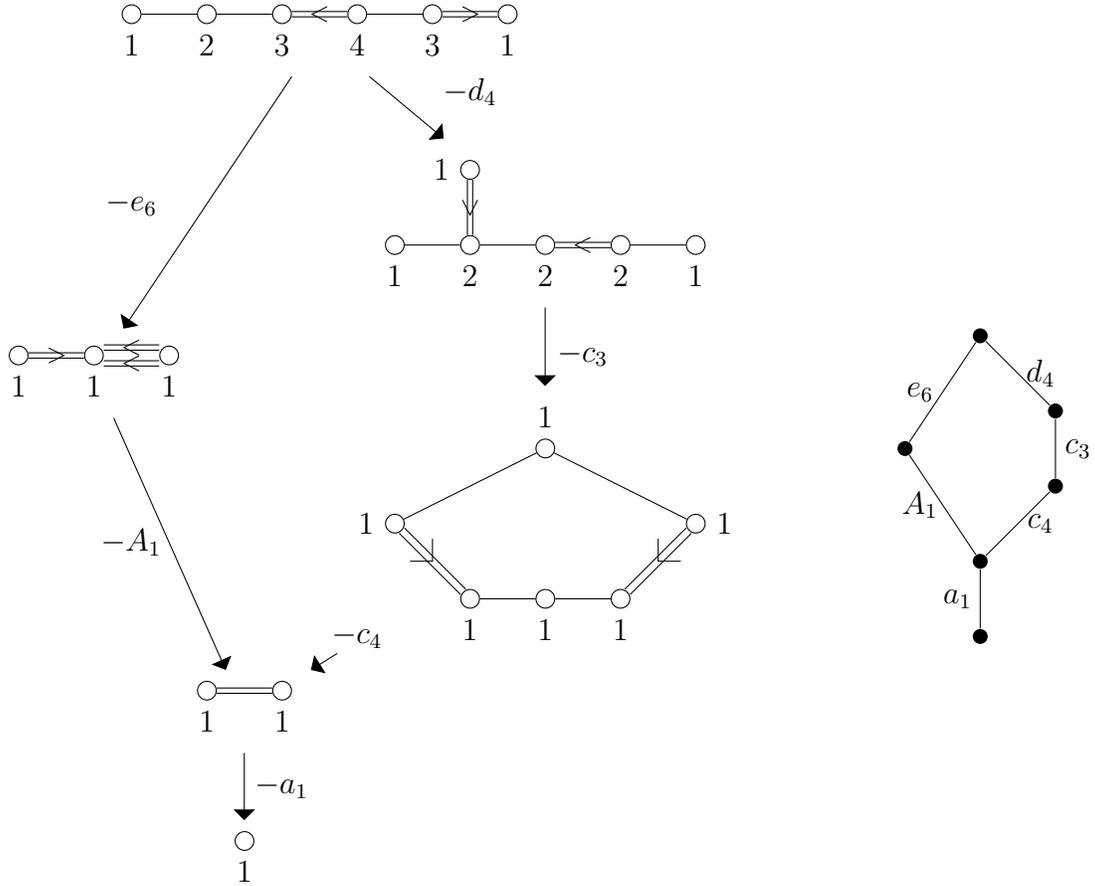

\paragraph{Example: Theory 34} The quiver subtraction for the magnetic quiver of theory \# 34 goes as shown in Figure \ref{fig:QS34}. 
The only slice visible at the bottom of the Hasse diagram is $a_1$ implying a non-abelian flavor symmetry of $A_1$. The full non-abelian flavor symmetry of theory 34, however, is $A_1A_3$. We comment on this behavior in the next paragraph.

\paragraph{Global Symmetry}

For several of our magnetic quivers quiver subtraction produces
\begin{equation}
	\begin{tikzpicture}
		\node[gauge,label=left:{$1$}] (l) at (0,0) {};
		\node[gauge,label=above:{$1$}] (1u) at (1,1) {};
		\node (u) at (2,1) {$\cdots$};
		\node[gauge,label=above:{$1$}] (2u) at (3,1) {};
		\node[gauge,label=below:{$1$}] (1d) at (1,-1) {};
		\node (d) at (2,-1) {$\cdots$};
		\node[gauge,label=below:{$1$}] (2d) at (3,-1) {};
		\node[gauge,label=right:{$1$}] (r) at (4,0) {};
		\draw (l)--(1u)--(u)--(2u)--(r) (1d)--(d)--(2d);
		
		\draw[transform canvas={xshift=1pt,yshift=1pt}] (l)--(1d);
		\draw[transform canvas={xshift=-1pt,yshift=-1pt}] (l)--(1d);
		\draw[transform canvas={xshift=1pt,yshift=-1pt}] (r)--(2d);
		\draw[transform canvas={xshift=-1pt,yshift=1pt}] (r)--(2d);
		
		\draw (0.5-0.3,-0.5)--(0.5,-0.5)--(0.5,-0.5+0.3);
		\draw (3.5,-0.5+0.3)--(3.5,-0.5)--(3.5+0.3,-0.5);
		
		\node at (0,-0.5) {$k$};
		\node at (4,-0.5) {$k$};
    	\draw[decorate,decoration={brace,amplitude=5pt}]
    (3.2,-1.6)--(0.8,-1.6) node [black,midway,yshift=-0.4cm] {$n$};	
    	\draw[decorate,decoration={brace,amplitude=5pt}]
    (0.8,1.6)--(3.2,1.6) node [black,midway,yshift=0.4cm] {$m$};	
	\end{tikzpicture}
	\label{eq:quiv_ns_loop}
\end{equation}
The CB of this quiver had global symmetry $A_mA_nU_1$ with highest weight generating function
\begin{equation}
	\mathrm{PE}[(1+\m_1\mu_m+\nu_1\nu_n)t^2+(q^{1}\mu_1\nu_n^k+q^{-1}\mu_m\nu_1^k)t^{k+1}-\mu_1\mu_m\nu_1^k\nu_n^kt^{2k+2}]
\end{equation}
For $k=1$ the global symmetry enhances to $A_{m+n+1}$, the CB is $a_{m+n+1}$, the HWG is still good. For $k=0$ the global symmetry enhances to $C_{m+n+1}$, the CB is freely generated, the HWG needs to be modified. From the quiver \eqref{eq:quiv_ns_loop} we compute the following Hasse diagram using quiver subtraction:

\begin{equation}
	\begin{tikzpicture}[scale=0.7]
		
		\node at (-8,-1) {$k=0$};
		\node[hasse,label=left:{$\mathbb{H}^{m+n+1}$}] at (-8,-4) {};

		\node at (-4,-1) {$k=1$};
		\node at (-4,-4) {$\begin{tikzpicture}
		\node[hasse] (a0) at (0,0) {};
		\node[hasse] (a2) at (0,2) {};
		\draw (a0)--(a2);
		\node at (-0.9,1) {$a_{m+n+1}$};
		\end{tikzpicture}$};
		
		\node at (1,-1) {$k>1$};
		\node at (4,-1) {$m=0$};
		\node at (8,-1) {$m\neq0$};
		\node at (1,-4) {$n=0$};
		\node at (4,-4) {$\begin{tikzpicture}
		\node[hasse] (a0) at (0,0) {};
		\node[hasse] (a2) at (0,2) {};
		\draw (a0)--(a2);
		\node at (-1,1) {$\bar{h}_{1,k}=A_{k}$};
		\end{tikzpicture}$};
		\node at (8,-4) {$\begin{tikzpicture}
		\node[hasse] (a0) at (0,0) {};
		\node[hasse] (a1) at (0,1) {};
		\node[hasse] (a2) at (0,2) {};
		\draw (a0)--(a1)--(a2);
		\node at (-0.4,0.5) {$a_{m}$};
		\node at (-1.1,1.5) {$h_{1,k}=A_{k-1}$};
		\end{tikzpicture}$};
		\node at (1,-8) {$n\neq0$};
		\node at (4,-8) {$\begin{tikzpicture}
		\node[hasse] (a0) at (0,0) {};
		\node[hasse] (a2) at (0,2) {};
		\draw (a0)--(a2);
		\node at (-0.7,1) {$\bar{h}_{n+1,k}$};
		\end{tikzpicture}$};
		\node at (8,-8) {$\begin{tikzpicture}
		\node[hasse] (a0) at (0,0) {};
		\node[hasse] (a1) at (0,1) {};
		\node[hasse] (a2) at (0,2) {};
		\draw (a0)--(a1)--(a2);
		\node at (-0.4,0.5) {$a_{m}$};
		\node at (-0.7,1.5) {$h_{n+1,k}$};
		\end{tikzpicture}$};
		\draw (0,-2)--(10,-2) (0,-6)--(10,-6) (2,0)--(2,-10) (6,0)--(6,-10);
	\end{tikzpicture}
\end{equation}

We notice, that for $m\neq0$ and $n\neq0$ we cannot read the full non-abelian part of the global symmetry from the bottom of the Hasse diagram produced from quiver subtraction. It is unclear, whether the Hasse diagram is nevertheless correct, and new methods are needed to check this. Several Hasse diagrams which share this property are known, as for example the Hasse diagram of instantons described below.

\paragraph{Two-instanton moduli spaces} The Hasse diagram of the moduli space of instantons is challenging to compute. Here we will present their diagram for 2 instantons, and give an explanation of it, deferring a more detailed discussion to an upcoming work \cite{InstantonFuture}. We will use brane physics to derive the diagram, since it is an intuitive route. SCFTs whose HBs are moduli spaces of instantons live on parallel D3 branes probing a (coincident) stack of $[p,q]7$-branes. Let the gauge group on the 7-branes be $G$ with Lie algebra $\mathfrak{g}$. The D3 branes have moduli: 1) transverse to the 7-branes, these are Coulomb directions; 2) inside the 7-branes, these are Higgs directions. To analyze the HB all D3s must be inside the 7-branes. A D3 inside the 7-branes has $1$ quaternionic position modulus, also it may bind together with the 7-branes leading to $(h^\vee_G-1)$ quaternionic moduli, containing the size modulus of the instanton. Abusing notation we refer to all these moduli simply as a size. Consider the case of two D3 branes, i.e.\ a rank-2 SCFT (provided a suitable set of $[p,q]$ $7$-branes was chosen). On a general point of the HB both D3s are far apart and have a size. We can now move to a lower leave by shrinking one D3 brane, this corresponds to a $\mathfrak{g}$ (minimal nilpotent orbit) transition. Since the D3 branes are identical, it doesn't matter which one we shrink, there is only one such transition. Now we can shrink the second D3, again leading to a $\mathfrak{g}$ transition. After the two D3 branes are shrunk, we can bring them together. Since the two branes are identical, this leads to a $\mathbb{C}^2/\mathbb{Z}_2$ transition. We are now left with a center of mass modulus in $\mathbb{H}$ which is smooth. The Hasse diagram obtained is:
\begin{equation}
    \begin{tikzpicture}
		\node[hasse] (a0) at (0,0) {};
		\node[hasse] (a1) at (0,1) {};
		\node[hasse] (a2) at (0,2) {};
		\node[hasse] (a3) at (0,3) {};
		\draw (a0)--(a1)--(a2)--(a3);
		\node at (-0.4,0.5) {$A_{1}$};
		\node at (-0.4,1.5) {$\mathfrak{g}$};
		\node at (-0.4,2.5) {$\mathfrak{g}$};
	\end{tikzpicture}
\end{equation}
Which can also be obtained from an extended quiver subtraction algorithm, which will be presented in \cite{InstantonFuture}. It is worth mentioning that the two-instanton moduli space, which, as mentioned above, realizes the HB of certain specific rank-two theories, was also indirectly analyzed in \cite{Beem:2019snk} via the explicit construction of the VOA for said theories. This analysis reproduces the HB relations from the strong generators of the VOA which are compatible with the Hilbert series of the two-instanton moduli space, computed for example in \cite{Gaiotto:2012uq,Hanany:2012dm,Keller:2012da,Cremonesi:2014xha}. Since the Hasse diagram above is obtained from a MQ which correctly reproduces these Hilbert series, we don't see any contradictions with our claim.

Just like in the case discussed before, the full (non-abelian) global symmetry of the HB cannot be read from the bottom of the Hasse diagram. The $\mathfrak{g}$ factor must be inferred from higher up the Hasse diagram. Another example of such an effect is the rank-1 $C_3A_1$ theory whose HB Hasse diagram was discussed in \cite{Bourget:2020asf}.

\paragraph{$\mathcal{N}=3$ theories} The magnetic quivers for the $G(3,1,2)$ and $G(4,1,2)$ $\mathcal{N}=3$ theories are
\begin{equation}
    \begin{tikzpicture}
            \node[gauge,label=below:{$1$}] (1) at (0,0) {};
            \node[gauge,label=below:{$2$}] (2) at (1,0) {};
            \draw[thick] (1)--(2);
            \draw (2) to [out=315,in=45,looseness=8] (2);
            \node at (0.5,0.3) {$n$};
            \node at (4,0) {for $G(n,1,2)$.};
    \end{tikzpicture}\;.
\end{equation}
Note that the integer $n$ on this quiver denotes the number of hypers, and there is \emph{no} arrow (the quiver is simply laced). 

We can use the quiver subtraction of \cite{InstantonFuture}, leading to the left line, or subtract $\raisebox{-0.5\height}{\scalebox{0.7}{
\begin{tikzpicture}
            \node[gauge,label=below:{$1$}] (1) at (0,0) {};
            \node[gauge,label=below:{$1$}] (2) at (1,0) {};
            \draw[thick] (1)--(2);
            \draw (2) to [out=315,in=45,looseness=8] (2);
            \node at (0.5,0.3) {$n$};
\end{tikzpicture}
}}
$, leading to the right line in:
\begin{equation}
    \begin{tikzpicture}
            \node[hasse] (1) at (0,0) {};
            \node[hasse] (2) at (-0.5,1) {};
            \node[hasse] (3) at (0.5,1) {};
            \node[hasse] (4) at (0,2) {};
            \node at (-0.7,0.5) {$A_{n-1}$};
            \node at (0.7,0.5) {$A_{n-1}$};
            \node at (-0.7,1.5) {$A_{1}$};
            \node at (0.7,1.5) {$A_{n-1}$};
            \draw (1)--(2)--(4)--(3)--(1);
    \end{tikzpicture}\;.
\end{equation}
This matches the explicit analysis in Appendix \ref{app:N=3}.

\acknowledgments We would like to thank Rudolph Kalveks for discussions and help with computations of bad quivers, and Behzat Ergun, Simone Giacomelli, Amihay Hanany and Zhenghao Zhong for discussions. AB is supported by the ERC Consolidator Grant 772408-Stringlandscape, and by the LabEx ENS-ICFP: ANR-10-LABX-0010/ANR-10-IDEX-0001-02 PSL*. JFG is supported by STFC grants ST/P000762/1 and ST/T000791/1. MM is supported by NSF grants PHY-1915093. GZ is supported in part by the ERC-STG grant 637844-HBQFTNCER, by the INFN and by the Simons Foundation grant 815892.

\clearpage 

\begin{table} 
\hspace*{-1cm}\begin{tabular}{cccccc} \toprule 
 \#   & $d_{HB}$ & $\mathfrak{f}$ & Quiver & UR & Method \\  \midrule 
 1 & $59+1$ & $[\mathfrak{e}_8]_{24} \times \mathfrak{su}(2)_{13}$ & \quiv{1} & 2  &  \ref{instanton} \ref{webs} \\
 2 & $46$ & $\mathfrak{so}(20)_{16}$ & \quiv{2}  &  2  & \ref{webs}  \\
 3 & $46$ & $[\mathfrak{e}_8]_{20}$ & \quiv{3}  & 2   &  \ref{webs}\\
 4 & $35+1$ & $[\mathfrak{e}_7]_{16} \times \mathfrak{su}(2)_{9}$ & \quiv{4}  & 2   &  \ref{webs} \ref{instanton}   \\
 5 & $30$ & $\mathfrak{su}(2)_{8} \times \mathfrak{so}(16)_{12}$ & \quiv{5}  & 2   & \ref{webs} \\
 6 & $26$ & $\mathfrak{su}(10)_{10}$ & \quiv{6}  & 2   & \ref{webs}  \ref{classS} \\
 7 & $23+1$ & $[\mathfrak{e}_6]_{12} \times \mathfrak{su}(2)_{7}$ & \quiv{7} &  2   &   \ref{instanton} \ref{webs}\\
 8 & $22$ & $\mathfrak{so}(14)_{10} \times \mathfrak{u}(1)$ & \quiv{8}  &  2  &  \ref{classS} \ref{webs}\\
 9 & $18$ & $\mathfrak{su}(2)_{6} \times \mathfrak{su}(8)_8$ & \quiv{9}   &  2  &  \ref{classS} \ref{webs}\\
 10 & $14$ & $\mathfrak{so}(12)_{8}$ & \quiv{10}  &  2   &  \ref{webs}  \\ \bottomrule 
\end{tabular}
    \caption{List of magnetic quivers (white nodes are balanced, black nodes are overbalanced). $d_{HB}$ is the quaternionic dimension of the Higgs branch, and $\mathfrak{f}$ is the flavor symmetry algebra. UR denotes the Unfolded Rank, i.e. the quaternionic dimension of the Higgs branch of the unfolded magnetic quiver. The last column points to the method used to derive the quiver.   }
    \label{tab:listMagneticQuivers1}
\end{table}

\begin{table} 
\begin{tabular}{cccccc} \toprule 
 \#   & $d_{HB}$ & $\mathfrak{f}$ & Quiver  &  UR   & Method \\  \midrule 
 11 & $12$ & $\mathfrak{so}(8)_{8} \times \mathfrak{su}(2)_5$ & \quiv{11}  &  2   &  \ref{instanton}, \ref{webs} \\
 12 & $10$ & $\mathfrak{u}(6)_6$ & \quiv{12} & 2 &   \ref{AD} \ref{webs}\\
 13 & $6$ & $\mathfrak{su}(2)_{4}^5$ &  \quiv{13} & 2 &  \ref{webs} \\
 14 & $6$ & $\mathfrak{su}(3)_{6} \times \mathfrak{su}(2)_4$ & \quiv{14} & 2 &   \ref{instanton}, \ref{webs} \\
 15 & $6$ & $\mathfrak{su}(5)_{5}$ & \quiv{15} & 2 & \ref{AD} \\
 16 & $4$ & $\mathfrak{su}(2)_{16/3} \times \mathfrak{su}(2)_{11/3}$ &  \quiv{16} &   2  & \ref{instanton}  \\
 17 & $2$ & $\mathfrak{su}(2)_{10/3} \times \mathfrak{u}(1)$  &  \quiv{17}  & 2  & \ref{AD} \\
 18 & $2$ & $\mathfrak{su}(2)_{17/5}$ & \quiv{18} &  2   &  \ref{instanton} \\
 19 & $1$ & $\mathfrak{su}(2)_{16/5}$ & \quiv{19} &  1   & \ref{AD} \\
 20 & $1$ & $\mathfrak{u}(1)$ &  \quiv{20} & 2  &  \ref{AD} \\
 21 & $0$ & $\emptyset$ &  \quiv{21} &  2   & \ref{AD} \\ \bottomrule
\end{tabular}
    \caption{List of magnetic quivers (continued).   }
    \label{tab:listMagneticQuivers2}
\end{table}

\begin{table} 
\hspace*{-2cm}\begin{tabular}{cccccc} \toprule 
 \#   & $d_{HB}$ & $\mathfrak{f}$ & Quiver  & UR   &  \\  \midrule
 22 & $22$ & $\mathfrak{sp}(12)_{8}$ &  \quiv{22}   & 3   & \ref{webs} \\ 
 23 & $20$ & $\mathfrak{sp}(4)_{7} \times \mathfrak{sp}(8)_{8} $ & \quiv{23}  & 3   &  \ref{webs} \\ 
 24 & $24$ & $\mathfrak{su}(2)_{7}^2 \times [\mathfrak{f}_4]_{12}$ &  \quiv{24}  &  3   & \ref{webs} \ref{Sfolds} \\ 
 25 & $12$ & $\mathfrak{su}(2)_{8} \times \mathfrak{sp}(8)_{6} $ &  \quiv{25}   &  3     & \ref{webs} \\  
 26 & $11$ & $\mathfrak{su}(2)_{5} \times \mathfrak{sp}(6)_{6} \times \mathfrak{u}(1)$  & \quiv{26}   & 3  &  \ref{classS} \\ 
 27 & $12$ &  $\mathfrak{su}(2)_{5}^2 \times \mathfrak{so}(7)_8$ & \quiv{27}   &  3  & \ref{webs} \ref{Sfolds}  \\
 28 & $16$ & $[\mathfrak{f}_4]_{10} \times \mathfrak{u}(1)$  & \quiv{28}   & 3    &  \ref{webs}\\
 29 & $7$ & $\mathfrak{sp}(6)_5 \times \mathfrak{u}(1)$  &  \quiv{29}  & 3   &  \ref{webs} \\
 30 & $6$ & $\mathfrak{su}(3)_6 \times \mathfrak{su}(2)_4^2$  &   \quiv{30} & 3   & \ref{webs} \ref{Sfolds}   \\
 31 & $3$ & $\mathfrak{sp}(4)_4$  & \quiv{31} & 2   &   \ref{webs} \\
\rowcolor{blue!07} 32 & $2$ &  $\mathfrak{su}(2)_3 \times \mathfrak{su}(2)_3$ &  \quiv{32}  &   & \ref{webs} \\ \bottomrule
\end{tabular}
    \caption{List of magnetic quivers (continued). Lines shaded in blue correspond to 4d $\mathcal{N}=4$ theories.   }
    \label{tab:listMagneticQuivers3}
\end{table}

\begin{table} 
\begin{tabular}{cccccc} \toprule 
 \#   & $d_{HB}$ &  $\mathfrak{f}$ & Quiver  &     &  \\  \midrule 
 33 & $23$ & $\mathfrak{su}(6)_{16} \times \mathfrak{su}(2)_{9}$   &  \quiv{33}  &  4   & \ref{webs}  \\
 34 & $13$ & $\mathfrak{su}(4)_{12} \times \mathfrak{su}(2)_{7} \times \mathfrak{u}(1)$  &   \quiv{34}   & 4    &  \ref{webs}\\
 35 & $11$ &  $\mathfrak{su}(3)_{10} \times \mathfrak{su}(3)_{10} \times \mathfrak{u}(1)$  &  \quiv{35}     &   4  & \ref{webs} \\
 36 & $8$ &  $\mathfrak{su}(3)_{10} \times \mathfrak{su}(2)_{6} \times \mathfrak{u}(1)$ & \quiv{36}   & 4  &  \ref{webs}  \\
 37 & $6$ &  $\mathfrak{su}(2)_{8} \times \mathfrak{su}(2)_{8} \times \mathfrak{u}(1)^2$ & \quiv{37}    & 4    & \ref{webs}  \\
 38 & $2$ &  $\mathfrak{u}(1)^2$ & \quiv{38}   &    3   &\ref{webs}  \\  \midrule 
 39 & 29 & $\mathfrak{sp}(14)_9$  &  \quiv{39}  &  4   &  \ref{webs} \\
 40 & 17 & $\mathfrak{su}(2)_8 \times \mathfrak{sp}(10)_7$  &  \quiv{40}  &  4    & \ref{webs} \\
 41 & 15 & $\mathfrak{su}(2)_5 \times \mathfrak{sp}(8)_7$  & \quiv{41} & ?  & \ref{classS} \\ 
 42 & 11 & $\mathfrak{sp}(8)_6 \times \mathfrak{u}(1)$  &  \quiv{42}   &  4  & \ref{webs}  \\
 43 & 6 &  $\mathfrak{sp}(6)_5$ &  \quiv{43}  &   3   &  \ref{webs} \\
  \bottomrule
\end{tabular}
    \caption{List of magnetic quivers (continued).   }
    \label{tab:listMagneticQuivers4}
\end{table}

\begin{table} 
\begin{tabular}{cccccc} \toprule 
 \#  & $d_{HB}$ &  $\mathfrak{f}$ & Quiver   &  UR  \\  \midrule 
 44 & $19$ &  $\mathfrak{su}(5)_{16}$ &  \quiv{44}   & 6   &  \ref{webs} \\
 45 & $6$  & $\mathfrak{su}(3)_{12} \times \mathfrak{u}(1) $  &  \quiv{45}  & 6    &  \ref{webs} \\
 46 & $3$ &  $\mathfrak{su}(2)_{10} \times \mathfrak{u}(1)$ &  \quiv{46}   &   5  &  \ref{webs} \\   \midrule 
 47 & $32$ &  $\mathfrak{sp}(12)_{11}$ & See Table \ref{tab:th47-49}  &  ?    &  \\  
 48 & $8$  & $\mathfrak{sp}(4)_{5} \times \mathfrak{so}(4)_8 $  & \quiv{48} & ?   &   \\ 
 49 & $14$ &  $\mathfrak{sp}(8)_{7}$ &   See Table \ref{tab:th47-49}  &   ?   &  \\  
 50  & $4$ &  $\mathfrak{sp}(4)_{13/3}$ &   \quiv{50}  &   ?   &  \\  \midrule 
 51 & 28  & $\mathfrak{sp}(8)_{13} \times \mathfrak{su}(2)_{26}$  &  \quiv{51}    &  4   &  \ref{webs} \ref{Sfolds} \\
 52 & 14 &  $\mathfrak{sp}(4)_{9} \times \mathfrak{su}(2)_{16} \times \mathfrak{su}(2)_{18}$ &  \quiv{52}    & 4    & \ref{webs} \ref{Sfolds}  \\
 53 & 7 & $\mathfrak{su}(2)_{7} \times \mathfrak{su}(2)_{14}  \times \mathfrak{u}(1)$  &  \quiv{53}   &  4   &   \ref{webs} \ref{Sfolds} \\
 54 & 6 & $\mathfrak{su}(2)_{6} \times \mathfrak{su}(2)_{8}$  &  \quiv{54}   &   5   &  \ref{webs} \ref{Sfolds}  \\
 55 & 2 & $\mathfrak{su}(2)_{5}$  &  \quiv{55}    & 5    & \ref{webs} \\
\rowcolor{blue!07} 56 & 2 & $\mathfrak{su}(2)_{10}$  &  \quiv{56}   &  ?   &  \ref{webs} \\     \bottomrule
\end{tabular}
    \caption{List of magnetic quivers (continued).  Lines shaded in blue correspond to 4d $\mathcal{N}=4$ theories.  }
    \label{tab:listMagneticQuivers5}
\end{table}

\begin{table} 
\begin{tabular}{cccccc} \toprule 
 \#   & $d_{HB}$ &  $\mathfrak{f}$ & Quiver   &    &   \\  \midrule 
 57 & 12 & $[\mathfrak{g}_2]_{8} \times \mathfrak{su}(2)_{14}$  & \quiv{57}    &  4    &  \ref{webs} \ref{Sfolds} \\
 58 & 4 & $\mathfrak{su}(2)_{16/3} \times \mathfrak{su}(2)_{10}$  &  \quiv{58}   &  4   &   \ref{webs} \ref{Sfolds} \\
 59 & 6 &  $[\mathfrak{g}_2]_{30/3}$ & \quiv{59}    & 4   &   \ref{webs} \\
\rowcolor{blue!07} 60 & 2 & $\mathfrak{su}(2)_{8}$  &  \quiv{60}    &   ?  &  \ref{webs} \\   \midrule 
 61 & 15 & $\mathfrak{su}(3)_{26} \times \mathfrak{u}(1)$  &  \quiv{61}    &   6  &  \ref{webs} \ref{Sfolds} \\
 62 & 5 & $\mathfrak{u}(1) \times \mathfrak{u}(1)$  &   \quiv{62}   &   6  & \ref{webs} \ref{Sfolds} \\
\rowcolor{green!07} 63 & 2 & $\mathfrak{u}(1)$  &  \quiv{63}   &   ?  &  \ref{webs} \\   \midrule 
 64 & 8 & $\mathfrak{su}(2)_{16} \times \mathfrak{u}(1)$  &   \quiv{64}  &  8    & \ref{webs} \ref{Sfolds}  \\
\rowcolor{green!07} 65 & 2 & $\mathfrak{u}(1)$  & \quiv{65}   &   ?   &   \ref{webs} \\     \midrule 
 66 & 10 & $\mathfrak{sp}(4)_{14} \times \mathfrak{su}(2)_8$  & \quiv{66}  &    &   \\  
 67 & 2 &  $\mathfrak{su}(2)_{14}$ &  \quiv{67}   &    &  \ref{webs} \\  
\rowcolor{blue!07} 68 & 2 & $\mathfrak{su}(2)_{14}$  &  \quiv{68}   &    & \ref{webs}  \\
 69 & 0 & $\emptyset$  &     &    &   \\    \bottomrule
\end{tabular}
    \caption{List of magnetic quivers (continued). Lines shaded in blue correspond to 4d $\mathcal{N}=4$ theories. Lines shaded in green correspond to 4d $\mathcal{N}=3$ theories.   }
    \label{tab:listMagneticQuivers6}
\end{table}

\begin{table} 
\hspace*{-1cm}\begin{tabular}{cccc} \toprule 
\# & $\mathfrak{f}$ & Quiver & \# free hypers  \\  \midrule
47 & $\mathfrak{sp}(12)_{11}$ & \begin{tabular}{c}
\quiv{471} \\ \quiv{472}
\end{tabular}  & \begin{tabular}{c}
5 \\ \\ \\ \\  4
\end{tabular}  \\  \midrule 
49 & $\mathfrak{sp}(8)_{7}$ & \begin{tabular}{c}
\quiv{491} \\ \quiv{492} \\ \quiv{493}
\end{tabular}  & \begin{tabular}{c}
4 \\ \\ \\ \\  4 \\ \\ \\ \\  3
\end{tabular}  \\
\bottomrule
\end{tabular}
    \caption{Magnetic quivers for theories 47 and 49. The 3d Coulomb branch of the quivers corresponds to the Higgs branch of the theories, plus a number of free hypermultiplets, given in the last column. Red nodes are underbalanced, white nodes are balanced and black nodes are overbalanced. The quivers can be derived from their $5d$ brane realization (see details in Appendix \ref{appendix5dWebs}) or from the associated class S description, which are related through the results of \cite{Zafrir:2016jpu}. There are actually several different webs, and as such class S theories, that realize these SCFTs, differing by the number of free hypers. This leads to multiple magnetic quivers for these theories. As there is no preferred choice between them, we have instead opted to list several of them. }
    \label{tab:th47-49}
\end{table}
\clearpage 
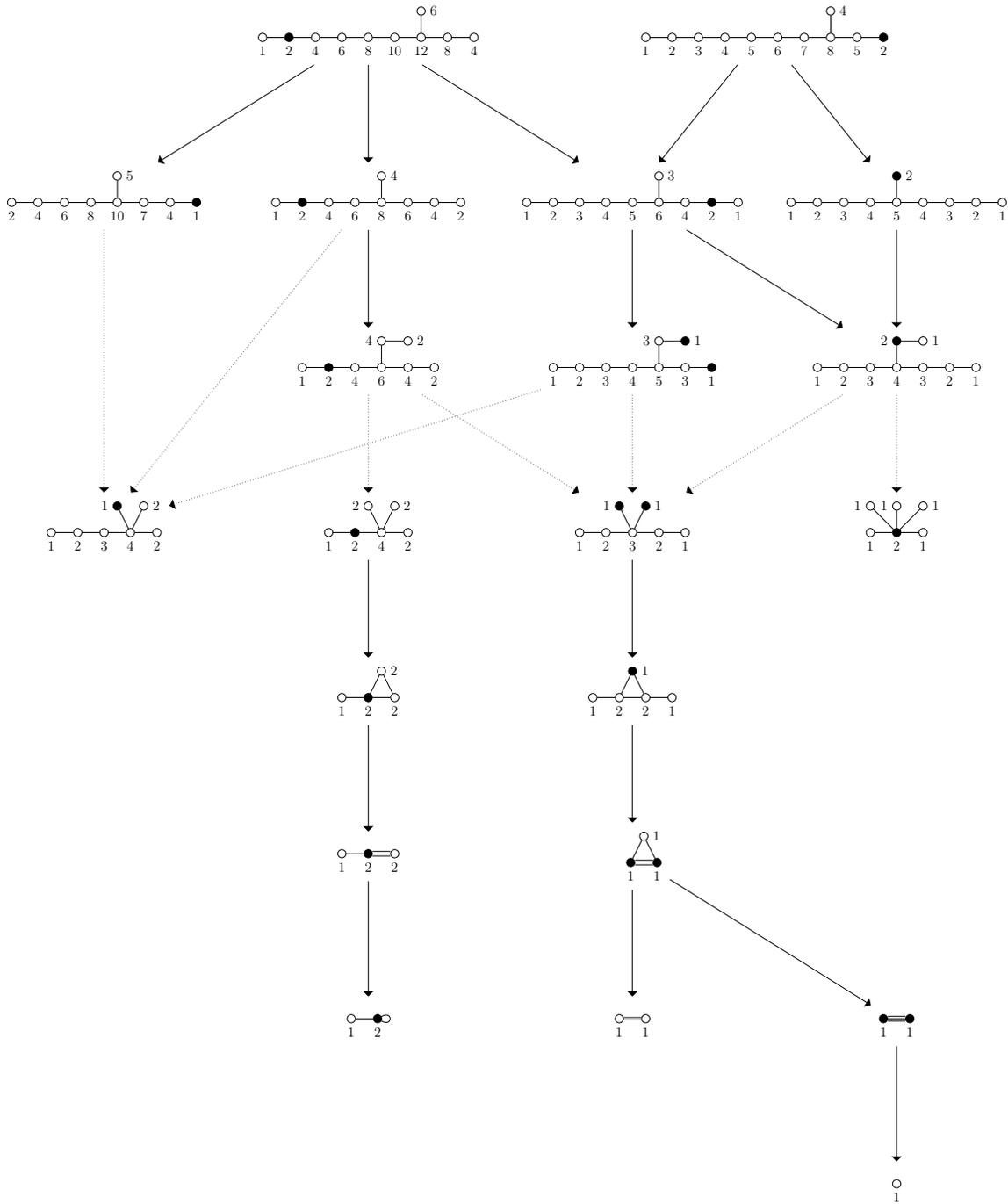
\begin{figure}
    \centering
\hspace*{-2cm}\scalebox{.5}{
\begin{tikzpicture}[xscale=8,yscale=5]
\node (t1) at (1,8) {\quiv{1}};
\node (t2) at (2.5,8) {\quiv{2}};
\node (t3) at (0,7) {\quiv{3}};
\node (t4) at (1,7) {\quiv{4}};
\node (t5) at (2,7) {\quiv{5}};
\node (t6) at (3,7) {\quiv{6}};
\node (t7) at (1,6) {\quiv{7}};
\node (t8) at (2,6) {\quiv{8}};
\node (t9) at (3,6) {\quiv{9}};
\node (t10) at (0,5) {\quiv{10}};
\node (t11) at (1,5) {\quiv{11}};
\node (t12) at (2,5) {\quiv{12}};
\node (t13) at (3,5) {\quiv{13}};
\node (t14) at (1,4) {\quiv{14}};
\node (t15) at (2,4) {\quiv{15}};
\node (t16) at (1,3) {\quiv{16}};
\node (t17) at (2,3) {\quiv{17}};
\node (t18) at (1,2) {\quiv{18}};
\node (t19) at (2,2) {\quiv{19}};
\node (t20) at (3,2) {\quiv{20}};
\node (t21) at (3,1) {\quiv{21}};
\draw[->] (t1)--(t3);
\draw[->] (t1)--(t4);
\draw[->] (t1)--(t5);
\draw[->] (t2)--(t5);
\draw[->] (t2)--(t6);
\draw[->] (t4)--(t7);
\draw[->] (t5)--(t8);
\draw[->] (t5)--(t9);
\draw[->] (t6)--(t9);
\draw[->] (t11)--(t14);
\draw[->] (t12)--(t15);
\draw[->] (t14)--(t16);
\draw[->] (t15)--(t17);
\draw[->] (t16)--(t18);
\draw[->] (t17)--(t19);
\draw[->] (t17)--(t20);
\draw[->] (t20)--(t21);
\draw[->,dotted] (t3)--(t10);
\draw[->,dotted] (t4)--(t10);
\draw[->,dotted] (t8)--(t10);
\draw[->,dotted] (t7)--(t11);
\draw[->,dotted] (t7)--(t12);
\draw[->,dotted] (t8)--(t12);
\draw[->,dotted] (t9)--(t12);
\draw[->,dotted] (t9)--(t13);
\end{tikzpicture}}
    \caption{The $\mathfrak{e}_8 - \mathfrak{so}(20)$ series}
    \label{fig:e8so20}
\end{figure}

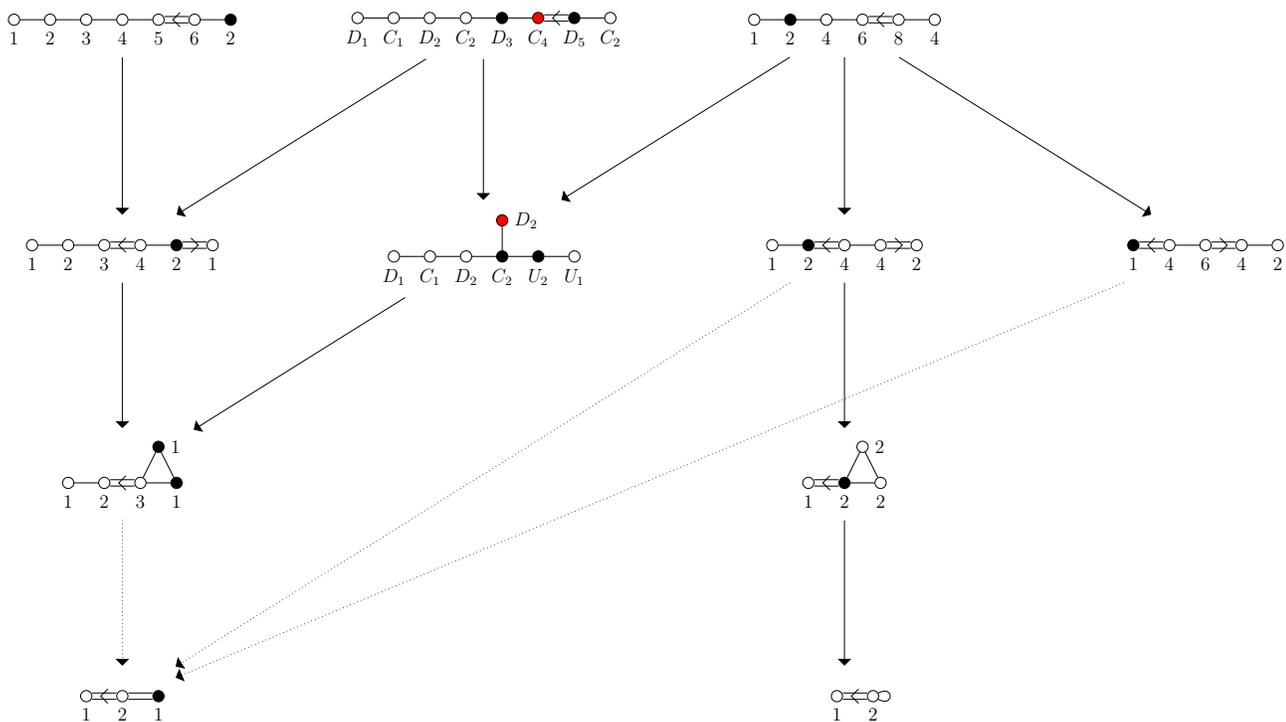
\begin{figure}
    \centering
\hspace*{-2cm}\scalebox{.6}{
\begin{tikzpicture}[xscale=8,yscale=5]
\node (t22) at (0,4) {\quiv{22}};
\node (t23) at (1,4) {\quiv{23}};
\node (t24) at (2,4) {\quiv{24}};
\node (t25) at (0,3) {\quiv{25}};
\node (t26) at (1,3) {\quiv{26}};
\node (t27) at (2,3) {\quiv{27}};
\node (t28) at (3,3) {\quiv{28}};
\node (t29) at (0,2) {\quiv{29}};
\node (t30) at (2,2) {\quiv{30}};
\node (t31) at (0,1) {\quiv{31}};
\node (t32) at (2,1) {\quiv{32}};
\draw[->] (t22)--(t25);
\draw[->] (t23)--(t25);
\draw[->] (t23)--(t26);
\draw[->] (t24)--(t26);
\draw[->] (t24)--(t27);
\draw[->] (t24)--(t28);
\draw[->] (t25)--(t29);
\draw[->] (t26)--(t29);
\draw[->] (t27)--(t30);
\draw[->] (t30)--(t32);
\draw[->,dotted] (t29)--(t31);
\draw[->,dotted] (t27)--(t31);
\draw[->,dotted] (t28)--(t31);
\end{tikzpicture}}
    \caption{The $\mathfrak{sp}(12) -\mathfrak{sp}(8) - \mathfrak{f}_4$ series. }
    \label{fig:sp12sp8f4}
\end{figure}

\begin{figure}
    \centering
\scalebox{.8}{
\begin{tikzpicture}[xscale=3,yscale=3]
\node (t33) at (0,3) {\quiv{33}};
\node (t34) at (-1,2) {\quiv{34}};
\node (t35) at (1,2) {\quiv{35}};
\node (t36) at (-1,1) {\quiv{36}};
\node (t37) at (1,1) {\quiv{37}};
\node (t38) at (0,0) {\quiv{38}};
\draw[->] (t33)--(t34);
\draw[->] (t33)--(t35);
\draw[->] (t34)--(t36);
\draw[->] (t34)--(t37);
\draw[->] (t35)--(t37);
\draw[->,dotted] (t36)--(t38);
\draw[->,dotted] (t37)--(t38);
\end{tikzpicture}}
    \caption{The $\mathfrak{su}(6)$ series. }
    \label{fig:su6}
\end{figure}
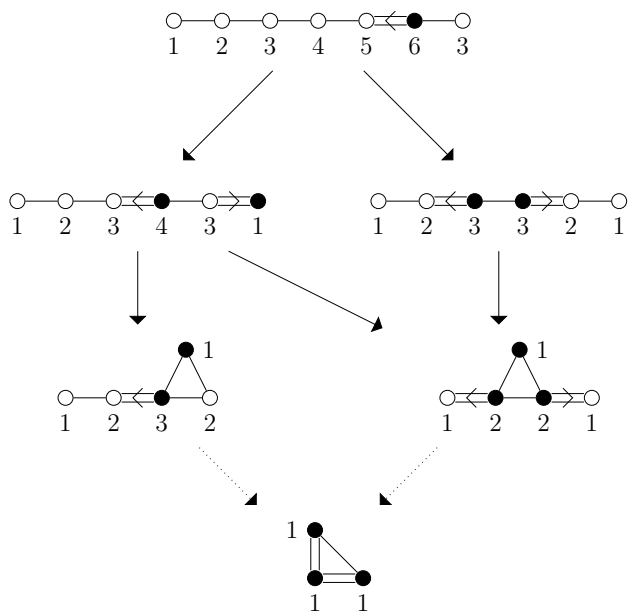

\begin{figure}
    \centering
\scalebox{.8}{
\begin{tikzpicture}[xscale=3,yscale=3]
\node (t39) at (0,3) {\quiv{39}};
\node (t40) at (-1,2) {\quiv{40}};
\node (t41) at (0,1) {\quiv{42}};
\node (t42) at (0,0) {\quiv{43}};
\node (t64) at (1,2) {\quiv{41}};
\draw[->] (t39)--(t40);
\draw[->] (t40)--(t41);
\draw[->] (t64)--(t41);
\draw[->,dotted] (t41)--(t42);
\end{tikzpicture}}
    \caption{The $\mathfrak{sp}(14)$ series. }
    \label{fig:sp14}
\end{figure}
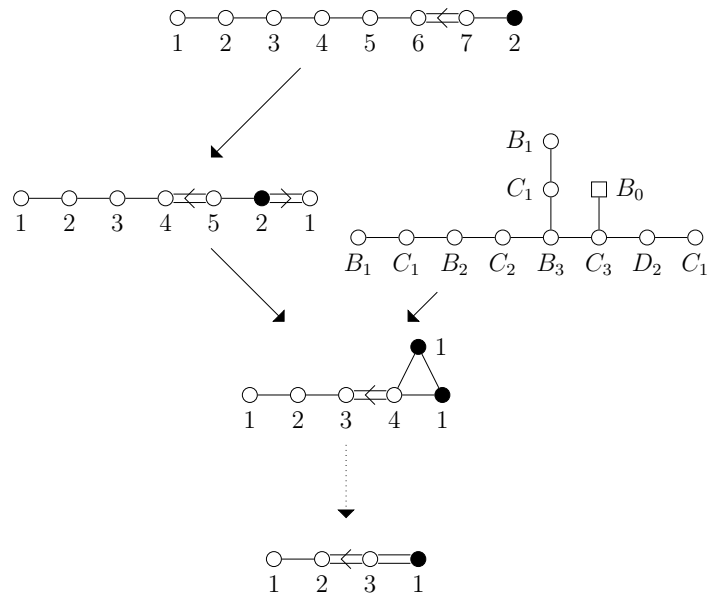

\begin{figure}
    \centering
\begin{tabular}{cc}
\scalebox{.8}{
\begin{tikzpicture}[xscale=3,yscale=3]
\node (t43) at (0,2) {\quiv{44}};
\node (t44) at (0,1) {\quiv{45}};
\node (t45) at (0,0) {\quiv{46}};
\draw[->] (t43)--(t44);
\draw[->] (t44)--(t45);
\end{tikzpicture}}
&
    \scalebox{.8}{
\begin{tikzpicture}[xscale=3,yscale=3]
\node (t49) at (-1,3) {\quiv{51}};
\node (t50) at (-1,2) {\quiv{52}};
\node (t51) at (-1,1) {\quiv{53}};
\node (t52) at (1,1) {\quiv{54}};
\node (t53) at (2,0) {\quiv{55}};
\node (t54) at (0,0) {\quiv{56}};
\draw[->] (t49)--(t50);
\draw[->] (t50)--(t51);
\draw[->] (t52)--(t54);
\draw[->] (t52)--(t53);
\draw[->,dotted] (t51)--(t54);
\end{tikzpicture}}
\end{tabular}    
\\ 
\begin{tabular}{ccc}
\scalebox{.8}{
\begin{tikzpicture}[xscale=3,yscale=3]
\node (t55) at (0,2) {\quiv{57}};
\node (t56) at (-1,1) {\quiv{58}};
\node (t57) at (1,1) {\quiv{59}};
\node (t58) at (-1,0) {\quiv{60}};
\draw[->] (t55)--(t56);
\draw[->] (t55)--(t57);
\draw[->] (t56)--(t58);
\end{tikzpicture}} & 
    \scalebox{.8}{
\begin{tikzpicture}[xscale=3,yscale=3]
\node (t59) at (0,2) {\quiv{61}};
\node (t60) at (0,1) {\quiv{62}};
\node (t61) at (0,0) {\quiv{63}};
\draw[->] (t59)--(t60);
\draw[->] (t60)--(t61);
\end{tikzpicture}} & 
\scalebox{.8}{
\begin{tikzpicture}[xscale=3,yscale=3]
\node (t62) at (0,1) {\quiv{64}};
\node (t63) at (0,0) {\quiv{65}};
\draw[->] (t62)--(t63);
\end{tikzpicture}}
\end{tabular}    
    \caption{The $\mathfrak{su}(5)$ series (top left), the $\mathfrak{sp}(8) - \mathfrak{su}(2)^2$ series (top right). The $\mathfrak{g}_2$ series (left), the $\mathfrak{su}(3)$ series (middle) and the $\mathfrak{su}(2)$ series (right). }
    \label{fig:su2}
\end{figure}
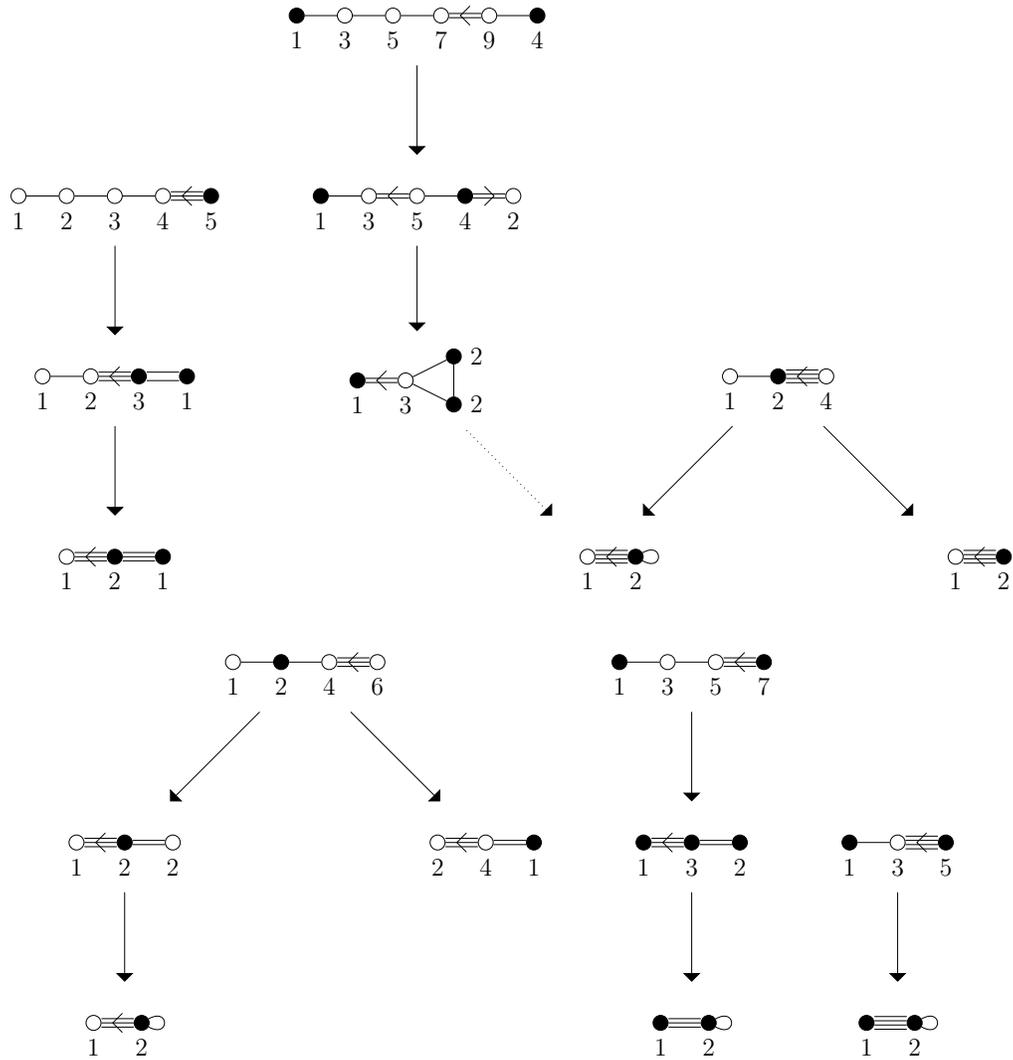
\clearpage 
\appendix

\section{5d webs}
\label{appendix5dWebs}

We list below the generalized toric polygons (GTPs) that encode brane webs for the 5d $\mathcal{N}=1$ theories listed in \cite{Martone:2021drm}. When needed, monodromies are realized on the GTPs, and the corresponding magnetic quiver is reported in Tables \ref{tab:listMagneticQuivers1}-\ref{tab:listMagneticQuivers6}.

\begin{enumerate}
\item[1. ] The theory is the 5d SCFT $\mathrm{SU}(3)_2 + 8F$. We use the standard polygon for $\mathrm{SU}(3)$ gauge theory with 7 fundamental hypers and one antisymmetric hyper from \cite{Zafrir:2015rga} (see the corresponding GTPs in \cite[Fig. 2]{vanBeest:2020civ}). Here and below, the arrow denotes a sequence of monodromies that are performed to make the polygon convex and describe the 5d SCFT point. 
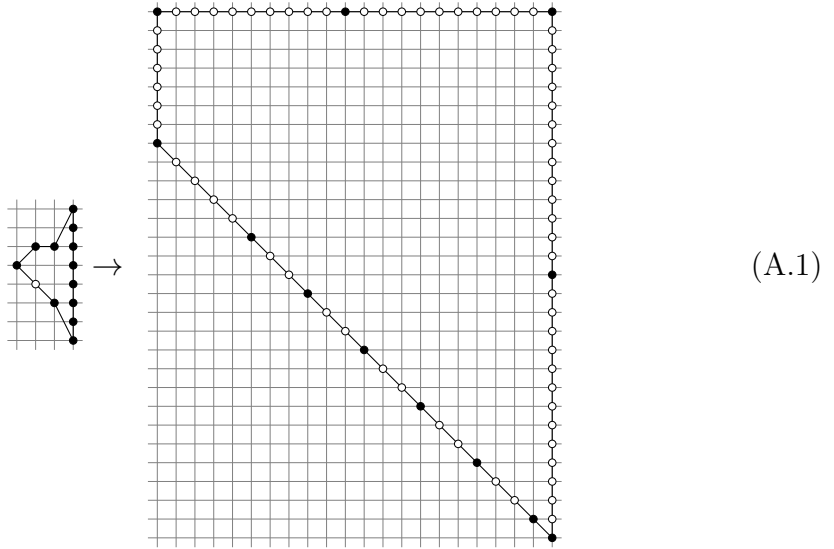
\begin{equation}
\raisebox{-.5\height}{ \scalebox{.5}{  
}}
\end{equation}
\item[2. ] The theory is the 5d SCFT $\mathrm{SU}(3)_{1/2} + 9F$. 
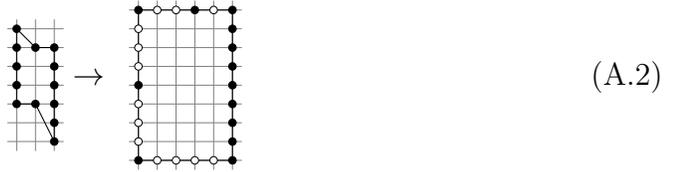
\begin{equation}
\raisebox{-.5\height}{ \scalebox{.5}{  
%
}}
\end{equation}
\item[3.] The theory is the the 5d SCFT quiver 
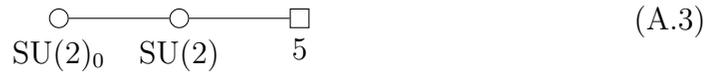
\begin{equation}
\label{quiverTh3}
\raisebox{-.5\height}{%
}
\end{equation}
and the polygons are 
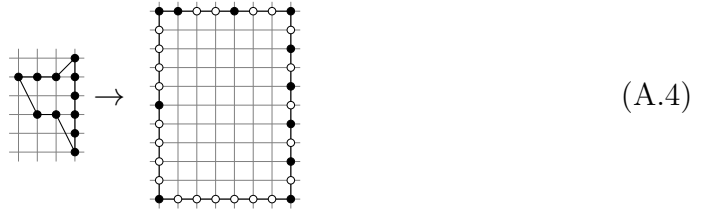
\begin{equation}
\raisebox{-.5\height}{ \scalebox{.5}{  
%
}}
\end{equation}
Note that it is important to correctly realize the theta angle for the leftmost SU(2) gauge group in the brane web. Changing this angle from 0 to $\pi$ corresponds to taking the polygon 
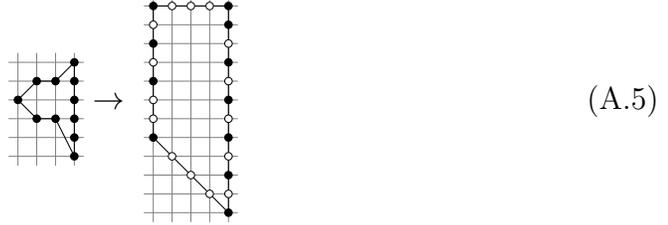
\begin{equation}
\raisebox{-.5\height}{ \scalebox{.5}{ \
%
}}
\end{equation}
instead. This corresponds to the next theory in this list (the $[\mathfrak{e}_7]_{16} \times \mathfrak{su}(2)_{9}$ theory). 
\item[4.]  The theory is the 5d SCFT $\mathrm{SU}(3)_{5/2} + 7F$. The polygon is 
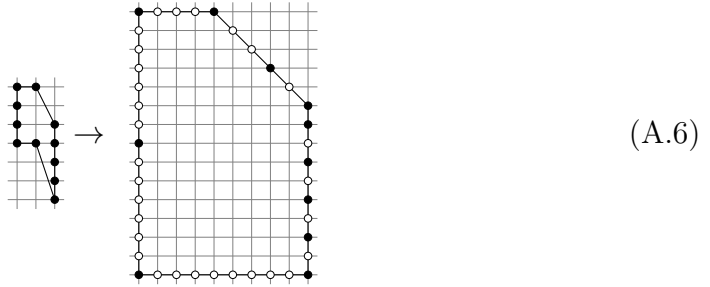
\begin{equation}
\raisebox{-.5\height}{ \scalebox{.5}{  
%
}}
\label{polygonTh4}
\end{equation}
As noted above, it can also be realized as an SCFT quiver theory, with quiver identical to (\ref{quiverTh3}) with the theta angle changed from 0 to $\pi$. One can check that the magnetic quiver that one gets from that description is the same as the one obtained from (\ref{polygonTh4}). From now on, we only pick one realization of a given SCFT to find the magnetic quiver, even when several are available. 
\item[5. ]  The theory is the 5d SCFT $\mathrm{SU}(3)_{1} + 8F$. The polygon is 
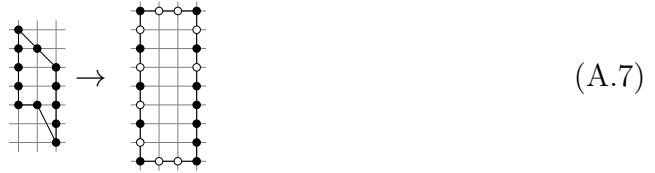
\begin{equation}
\raisebox{-.5\height}{ \scalebox{.5}{  
%
}}
\end{equation}
\item[6.]  The theory is the 5d SCFT $\mathrm{SU}(3)_{0} + 8F$. The polygon is 
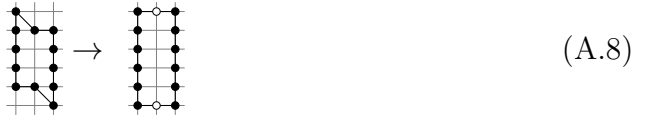
\begin{equation}
\raisebox{-.5\height}{ \scalebox{.5}{  
%
}}
\end{equation}
\item[7. ]  The theory is the 5d SCFT $\mathrm{SU}(3)_{3} + 6F$. The polygon is 
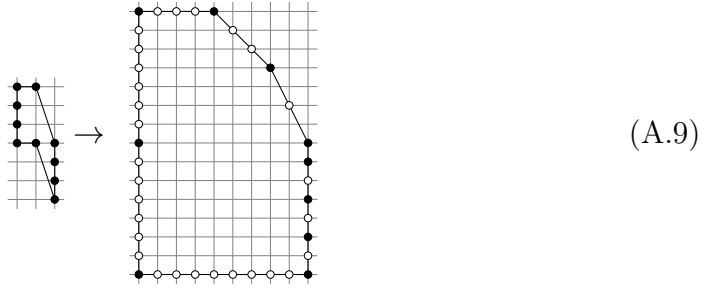
\begin{equation}
\raisebox{-.5\height}{ \scalebox{.5}{  
%
}}
\end{equation}
\item[8. ] The theory is the 5d SCFT $\mathrm{SU}(3)_{3/2} + 7F$. The polygon is 
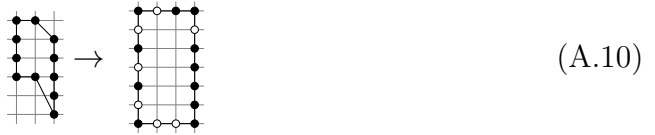
\begin{equation}
\raisebox{-.5\height}{ \scalebox{.5}{  
%
}}
\end{equation}
\item[9. ] The theory is the 5d SCFT $\mathrm{SU}(3)_{1/2} + 7F$. The polygon is 
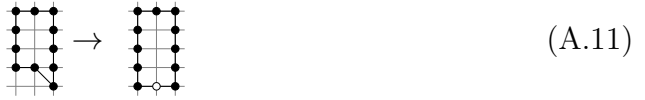
\begin{equation}
\raisebox{-.5\height}{ \scalebox{.5}{  
%
}}
\end{equation}
\item[10. ] The theory is the gauge theory $\mathrm{USp}(4) + 6F$. We use the construction from \cite[Fig. 6]{Bergman:2015dpa}. The line inside the polygon indicates a resolution, and corresponds to the fact that we consider the gauge theory phase, not the SCFT phase.  The same comments apply to the next few theories in this list. The polygon is 
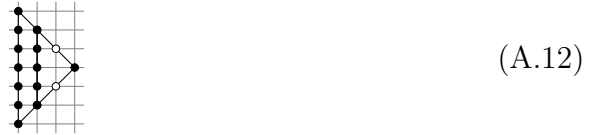
\begin{equation}
\raisebox{-.5\height}{ \scalebox{.5}{  
%
}}
\end{equation}
\item[11. ] The theory is the gauge theory $\mathrm{USp}(4) + 1 \mathrm{AS} + 4F$. We use the construction of \cite[Fig. 46]{Zafrir:2015rga}. The polygon is 
\begin{equation}
\raisebox{-.5\height}{ \scalebox{.5}{  
%
}}
\end{equation}
\item[12. ] The theory is the gauge theory $\mathrm{SU}(3) + 6F$. The polygon is 
\begin{equation}
\raisebox{-.5\height}{ \scalebox{.5}{  
%
}}
\end{equation}
\item[13. ] The theory is the quiver gauge theory
\begin{equation}
\raisebox{-.5\height}{%
}
\end{equation}
The polygon is (we apply a 90 degrees rotation): 
\begin{equation}
\raisebox{-.5\height}{ \scalebox{.5}{  
%
}}
\end{equation}
\item[22. ] The theory is the 5d SCFT $\mathrm{SU}(4)_{0} + 10F$, with a $\mathbb{Z}_2$ twist. We indicate this twist by the symbol $/ \mathbb{Z}_2$ on the GTP obtained after monodromies. The magnetic quiver is then computed using the rules spelled out in \cite{Bourget:2020mez}. The same notation is adpoted in the rest of this list. The polygon is 
\begin{equation}
\raisebox{-.5\height}{ \scalebox{.5}{  
%
}}  / \mathbb{Z}_2
\end{equation}
\item[23. ] The theory is the 5d SCFT $\mathrm{SU}(4)_{0} + 1 \mathrm{AS} + 8F$, with a $\mathbb{Z}_2$ twist. 
\begin{equation}
\raisebox{-.5\height}{ \scalebox{.5}{  
%
}}
\end{equation}
Here the resulting GTP is not $\mathbb{Z}_2$ symmetric. Of course, the sequence of monodromies that lead from the left hand side GTP to a convex one is not unique, but we could not find any such sequence that would yield a symmetric GTP. We conjecture that such a sequence does not exist. The magnetic quiver for the rightmost GTP above is 
\begin{equation}
 \raisebox{-.5\height}{%
}
\end{equation}
and correspondingly it is not $\mathbb{Z}_2$ symmetric, preventing folding. 
For that reason we turn to orthosymplectic quivers for that theory. 

To construct a brane web for the theory Spin(6) with one vector and 8 spinors, one uses the construction of \cite[Fig. 26]{Zafrir:2015ftn}. Because of the charges of the O5$^+$ and O5$^-$ planes, we indicate these orientifolds by coloring in red the five-branes that end on them, with alternating slope $+2$ and $-2$ for charge conservation:  
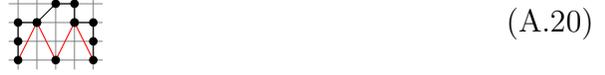
\begin{equation}
\raisebox{-.5\height}{ \scalebox{.5}{  
%
}}
\end{equation}
The triangles on the left and the right produce 4 spinors each, while the central pentagon realizes the Spin(6) theory with one vector. The monodromies are constrained not to affect the orientifold (below the red line), and can be performed as follows: 
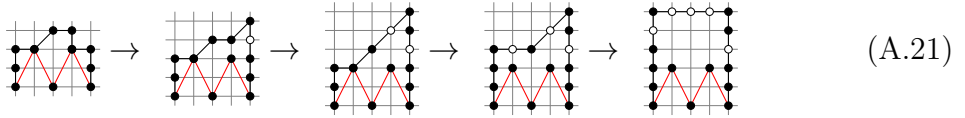
\begin{equation}
\label{webTh23}
\raisebox{-.5\height}{ \scalebox{.5}{  
%
}}
\end{equation}
It is remarkable that in this description, the brane web does have a $\mathbb{Z}_2$ symmetry. 
On the resulting configuration one can read a magnetic quiver following \cite{Zafrir:2015ftn,Bourget:2020gzi}. This gives 
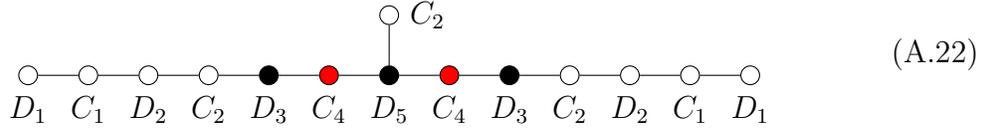
\begin{equation}
\raisebox{-.5\height}{%
}
\end{equation}
This quiver can be folded, giving finally the magnetic quiver shown in Table \ref{tab:listMagneticQuivers3}. 
\item[24. ] The theory is the 5d SCFT $\mathrm{SU}(4)_{0} + 2 \mathrm{AS} + 6F$, with a $\mathbb{Z}_2$ twist. Here the GTP is immediately convex and $\mathbb{Z}_2$ symmetric: 
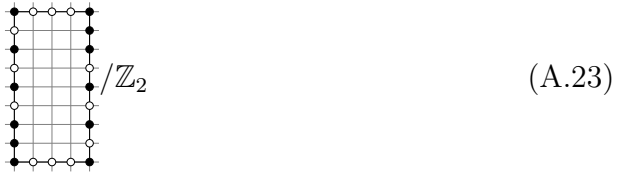
\begin{equation}
\raisebox{-.5\height}{ \scalebox{.5}{  
%
}} / \mathbb{Z}_2
\end{equation}
\item[25. ] The theory is the 5d SCFT $\mathrm{SU}(4)_{0}+ 8F$, with a $\mathbb{Z}_2$ twist. The polygon is 
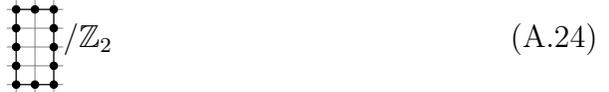
\begin{equation}
\raisebox{-.5\height}{ \scalebox{.5}{  
%
}} / \mathbb{Z}_2
\end{equation}
\item[26. ] The theory is the 5d SCFT $\mathrm{SU}(4)_{0} + 1 \mathrm{AS} + 6F$, with a $\mathbb{Z}_2$ twist. 
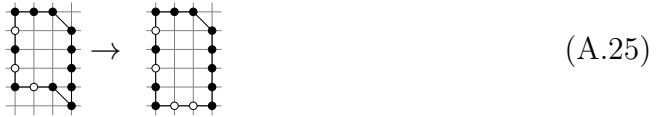
\begin{equation}
\raisebox{-.5\height}{ \scalebox{.5}{  
%
}}
\end{equation}
This is not $\mathbb{Z}_2$ symmetric, and the same comments made for item \# 23 in this list apply. The magnetic quiver for the rightmost GTP is 
\begin{equation}
     \raisebox{-.5\height}{%
}
\end{equation}
and correspondingly it is not $\mathbb{Z}_2$ symmetric, preventing folding. 
\item[27. ] The theory is the 5d SCFT $\mathrm{SU}(4)_{0} + 2 \mathrm{AS} + 4F$, with a $\mathbb{Z}_2$ twist. The polygon is 
\begin{equation}
\raisebox{-.5\height}{ \scalebox{.5}{  
%
}} / \mathbb{Z}_2
\end{equation}
\item[28. ] The theory is the 5d SCFT quiver 
\begin{equation}
\raisebox{-.5\height}{%
}
\end{equation}
with a $\mathbb{Z}_2$ twist. The polygon is 
\begin{equation}
\raisebox{-.5\height}{ \scalebox{.5}{  
%
}} / \mathbb{Z}_2
\end{equation}
\item[29. ] The theory is the 5d SCFT $\mathrm{SU}(4)_{0} + 6F$, with a $\mathbb{Z}_2$ twist. The polygon is 
\begin{equation}
\raisebox{-.5\height}{ \scalebox{.5}{  
%
}} / \mathbb{Z}_2
\end{equation}
\item[30. ] The theory is the 5d SCFT $\mathrm{SU}(4)_{0} + 2 \mathrm{AS} + 2F$, with a $\mathbb{Z}_2$ twist. The polygon is 
\begin{equation}
\raisebox{-.5\height}{ \scalebox{.5}{  
%
}} / \mathbb{Z}_2
\end{equation}
\item[31. ] The theory is the gauge theory $\mathrm{SU}(4)_{0} + 4F$, with a $\mathbb{Z}_2$ twist. The polygon is 
\begin{equation}
\raisebox{-.5\height}{ \scalebox{.5}{  
%
}} / \mathbb{Z}_2
\end{equation}
\item[32. ] The theory is the gauge theory $\mathrm{SU}(4)_{0} + 2 \mathrm{AS}$, with a $\mathbb{Z}_2$ twist. The polygon is 
\begin{equation}
\raisebox{-.5\height}{ \scalebox{.5}{  
%
}} / \mathbb{Z}_2
\end{equation}
\item[33. ] The theory is the 5d SCFT quiver 
\begin{equation}
\raisebox{-.5\height}{%
}
\end{equation}
with a $\mathbb{Z}_2$ twist. The polygon is 
\begin{equation}
\raisebox{-.5\height}{ \scalebox{.5}{  
%
}} / \mathbb{Z}_2
\end{equation} 
\item[34. ] The theory is the 5d SCFT quiver 
\begin{equation}
\raisebox{-.5\height}{%
}
\end{equation}
with a $\mathbb{Z}_2$ twist. The polygon is 
\begin{equation}
\raisebox{-.5\height}{ \scalebox{.5}{  
%
}} / \mathbb{Z}_2
\end{equation}
\item[35. ] The theory is the 5d SCFT quiver 
\begin{equation}
\raisebox{-.5\height}{%
}
\end{equation}
with a $\mathbb{Z}_2$ twist. The polygon is 
\begin{equation}
\raisebox{-.5\height}{ \scalebox{.5}{  
%
}} / \mathbb{Z}_2
\end{equation}
\item[36. ] The theory is the 5d SCFT quiver 
\begin{equation}
\raisebox{-.5\height}{%
}
\end{equation}
with a $\mathbb{Z}_2$ twist. The polygon is 
\begin{equation}
\raisebox{-.5\height}{ \scalebox{.5}{  
%
}} / \mathbb{Z}_2
\end{equation}
\item[37. ] The theory is the 5d SCFT quiver 
\begin{equation}
\raisebox{-.5\height}{%
}
\end{equation}
with a $\mathbb{Z}_2$ twist. The polygon is 
\begin{equation}
\raisebox{-.5\height}{ \scalebox{.5}{  
%
}} / \mathbb{Z}_2
\end{equation}
\item[38. ] The theory is the quiver gauge theory
\begin{equation}
\raisebox{-.5\height}{%
}
\end{equation}
with a $\mathbb{Z}_2$ twist. The Chern-Simons level $k$ can take any integer value, this does not affect the Higgs branch. The polygon is 
\begin{equation}
\raisebox{-.5\height}{ \scalebox{.5}{  
%
}} / \mathbb{Z}_2
\end{equation}
\item[39. ] The theory is the 5d SCFT $\mathrm{SU}(5)_{0} + 12F$. 
\begin{equation}
\raisebox{-.5\height}{ \scalebox{.5}{  
%
}} / \mathbb{Z}_2
\end{equation}
\item[40. ] The theory is the 5d SCFT $\mathrm{SU}(5)_{0} + 10F$ with a $\mathbb{Z}_2$ twist.  
\begin{equation}
\raisebox{-.5\height}{ \scalebox{.5}{  
%
}} / \mathbb{Z}_2
\end{equation}
\item[41. ] No 5d SCFT construction is known. 
\item[42. ] The theory is the 5d SCFT $\mathrm{SU}(5)_{0} + 8F$ with a $\mathbb{Z}_2$ twist. 
\begin{equation}
\raisebox{-.5\height}{ \scalebox{.5}{  
%
}} / \mathbb{Z}_2
\end{equation}
\item[43. ] The theory is the gauge theory $\mathrm{SU}(5)_{0} + 6F$ with a $\mathbb{Z}_2$ twist. 
\begin{equation}
\raisebox{-.5\height}{ \scalebox{.5}{  
%
}} / \mathbb{Z}_2
\end{equation}
\item[44. ]  This is the $T_5$ 5d SCFT with a $\mathbb{Z}_3$ twist. 
\begin{equation}
\raisebox{-.5\height}{ \scalebox{.5}{  
%
}} / \mathbb{Z}_3
\end{equation}
\item[45. ] The theory is the 5d quiver SCFT
\begin{equation}
\label{quiverth45}
\raisebox{-.5\height}{%
}}/ \mathbb{Z}_3
\end{equation}
\item[46. ] No known Lagrangian is known for this 5d SCFT. However the GTP can be obtain by decoupling one more hyper on the right of quiver (\ref{quiverth45}) and imposing $\mathbb{Z}_3$ symmetry. 
\begin{equation}
\raisebox{-.5\height}{ \scalebox{.5}{  
%
}}/ \mathbb{Z}_3
\end{equation}
\item[47. ]   The theory is the 5d SCFT corresponding to $G_2$ with five fundamentals. This can be realized with the following brane web, where the red line corresponds to five-branes which connect to $\widetilde{\mathrm{O5}}{}^{\pm}$ planes. See the explanations for theory \# 49 below on how this web is obtained. 
\begin{equation}
\raisebox{-.5\height}{ \scalebox{.5}{  
%
}}
\end{equation}
The magnetic quiver that one gets from that construction is the second one in Table \ref{tab:th47-49}. 
\item[49. ] The theory is the gauge theory $G_2$ with four fundamentals. We briefly review the construction of brane webs for $G_2$ theories from \cite{Hayashi:2018bkd}. The starting point is the pure Spin(7) theory, which is described by 
\begin{equation}
\raisebox{-.5\height}{ \scalebox{.5}{  
%
}}
\end{equation}
where the red lines represent five-branes ending on an $\widetilde{\mathrm{O5}} {}^{\pm}$ plane (here an $\widetilde{\mathrm{O5}} {}^{+}$ extends at infinity, while an $\widetilde{\mathrm{O5}} {}^{-}$ stretches between the two five-branes). We recall that a half-monodromy is stuck on the $\widetilde{\mathrm{O5}} {}^{\pm}$, which is responsible for the asymetry of the diagram. One then adds matter as described in \cite{Zafrir:2015ftn}; for instance two spinors and three vectors give  
\begin{equation}
\raisebox{-.5\height}{ \scalebox{.5}{  
%
}}
\end{equation}
Finally, one higgses one spinor to get $G_2$ with four fundamentals (and three singlets, coming from the three Spin(7) vectors), yielding 
\begin{equation}
\raisebox{-.5\height}{ \scalebox{.5}{  
%
}}
\end{equation}
where the internal line has been restablished to indicate that the $G_2$ gauge coupling should have finite value. 

Another method is to proceed as follows: it is known that the 5d SCFT corresponding at low energies to $G_2$ with four fundamentals is the same as the 5d SCFT corresponding to the gauge theory $\mathrm{USp}(4) + 2AS + 2F$. Therefore, we start with a construction of the latter (left polygon below), then perform the necessary monodromies to reach the UV fixed point (middle polygon), and finally partially resolve to get the $G_2$ gauge theory (right). 
\begin{equation}
\raisebox{-.5\height}{ \scalebox{.5}{  
}}
\end{equation}
Of course we get the same result as with the other method. 
The magnetic quiver that one gets from that construction is the last one in Table \ref{tab:th47-49}. 
\item[51. ]  The theory is the 5d SCFT $\mathrm{SU}(5)_{0} + 2 \mathrm{AS} + 6F$ with a $\mathbb{Z}_2$ twist. 
\begin{equation}
\raisebox{-.5\height}{ \scalebox{.5}{  
%
}}  / \mathbb{Z}_2
\end{equation}
\item[52. ] The theory is the 5d SCFT $\mathrm{SU}(5)_{0} + 2 \mathrm{AS} + 4F$ with a $\mathbb{Z}_2$ twist. 
\begin{equation}
\raisebox{-.5\height}{ \scalebox{.5}{  
%
}}  / \mathbb{Z}_2
\end{equation}
\item[53. ] The theory is the 5d SCFT $\mathrm{SU}(5)_{0} + 2 \mathrm{AS} + 2F$ with a $\mathbb{Z}_2$ twist.
\begin{equation}
\raisebox{-.5\height}{ \scalebox{.5}{  
%
}}  / \mathbb{Z}_2
\end{equation}
\item[54. ]  
The theory is the 5d SCFT quiver 
\begin{equation}
\raisebox{-.5\height}{%
}} / \mathbb{Z}_4
\end{equation}
\item[55. ] 
The theory is the 5d SCFT quiver 
\begin{equation}
\raisebox{-.5\height}{%
}}  / \mathbb{Z}_4
\end{equation}
Note that in the above polygon, the $\theta$-angle for the middle $\mathrm{SU}(2)$ gauge group is indeed $\pi$. This can be checked as follows: the configuration is symmetric, and one can decouple the six hypers by giving a large positive mass to three hypers and a large negative mass to three hypers, i.e. an odd number of negative masses. 
\item[56. ]  
The theory is the 5d quiver gauge theory
\begin{equation}
\raisebox{-.5\height}{%
}} / \mathbb{Z}_4
\end{equation}
\item[57. ]  
The theory is the 5d SCFT quiver 
\begin{equation}
\raisebox{-.5\height}{%
}
\end{equation}
with a $\mathbb{Z}_3$ twist. We construct separately the GTPs for the SU(4) part and for the SU(2) part, and them assemble them before making the result convex using monodromies: 
\begin{equation}
\raisebox{-.5\height}{ \scalebox{.5}{  
%
}} / \mathbb{Z}_3
\end{equation}
\item[58. ] 
The theory is the 5d SCFT quiver 
\begin{equation}
\raisebox{-.5\height}{%
}
\end{equation}
with a $\mathbb{Z}_3$ twist. We proceed as before, and the last arrow below is an $\mathrm{SL}(2 , \mathbb{Z})$ transformation which makes the $\mathbb{Z}_3$ symmetry apparent. 
\begin{equation}
\raisebox{-.5\height}{ \scalebox{.5}{  
%
}} / \mathbb{Z}_3
\end{equation}
\item[59. ] 
The theory is the 5d SCFT quiver 
\begin{equation}
\raisebox{-.5\height}{%
}
\end{equation}
with a $\mathbb{Z}_3$ twist. To construct the GTP, one can start from the theory 
\begin{equation}
\raisebox{-.5\height}{%
}
\end{equation}
which is realized by 
\begin{equation}
\raisebox{-.5\height}{ \scalebox{.5}{  
%
}}
\end{equation}
and add the flavor, giving 
\begin{equation}
\raisebox{-.5\height}{ \scalebox{.5}{  
%
}} / \mathbb{Z}_3
\end{equation}
\item[61. ]
The theory is the 5d SCFT quiver 
\begin{equation}
\raisebox{-.5\height}{%
}} / \mathbb{Z}_3
\end{equation}
\item[62. ]
The theory is the 5d SCFT quiver 
\begin{equation}
\raisebox{-.5\height}{%
}} / \mathbb{Z}_3
\end{equation}
\item[64. ]
The theory is the 5d SCFT quiver 
\begin{equation}
\raisebox{-.5\height}{%
}} / \mathbb{Z}_4
\end{equation}
\item[65. ] The theory is the 5d SCFT 
\begin{equation}
\raisebox{-.5\height}{%
}} / \mathbb{Z}_4
\end{equation}
\end{enumerate}

\section{Hasse diagrams}
\label{appendixHasse}

The Hasse diagrams that are deduced from quiver subtraction -- only for unitary quivers -- are shown in Tables \ref{fig:HasseDiagramsTable1}, \ref{fig:HasseDiagramsTable2}, \ref{fig:HasseDiagramsTable3}, \ref{fig:HasseDiagramsTable4}. We use the following color code to state the cross-checks that were performed on these diagrams: 
\begin{itemize}
    \item White: the Hasse diagram can be obtained with quiver subtraction, and with other methods \cite{Martone:2021ixp}, and the results agree. 
    \item Green: the Hasse diagram can only be obtained with quiver subtraction. 
    \item Red: the Hasse diagram can not be obtained using quiver subtraction, but it is computed in \cite{Martone:2021ixp}. In those cases, we don't draw any diagram and refer to \cite{Martone:2021ixp}. 
\end{itemize}

\input{hasseDiagrams}

\clearpage

\section{\texorpdfstring{The Hall-Littlewood index of the $USp(4)\times USp(8)$ $4d$ SCFT}{The Hall-Littlewood index of the USp(4)xUSp(8) 4d SCFT}}
\label{appendixHLindex}

As we previously mentioned, there are indications that the $USp(4)\times USp(8)$ $4d$ SCFT can be generated by the twisted compactification of the $5d$ SCFT UV completing the $5d$ gauge theory $SU(4)_0+1AS+8F$, see \cite{Martone:2021drm} for the full discussion. The latter has a brane web realization, using an $O5^-$ plane, from which we can generate an orthosymplectic magnetic quiver. This quiver manifests a $\mathbb{Z}_2$ symmetry, which is tempting to identify with the symmetry that we twist by. If so we can get a magnetic quiver for the $USp(4)\times USp(8)$ $4d$ SCFT by folding the quiver, which is one of those we proposed for this theory. Indeed the dimension of the Coulomb branch of this quiver fits the dimension of the Higgs branch of the $USp(4)\times USp(8)$ $4d$ SCFT. However, the resulting quiver has a bad node making further checks difficult.

Nevertheless, there is an interesting check that we can make. In addition to the magnetic quiver, we can associate with the pre-folding magnetic quiver a class S theory, whose $3d$ reduction is the mirror of the magnetic quiver. We can then compute the Hilbert series of this magnetic quiver from the Hall-Littlewood index of the associated class S theories\footnote{These usually coincide for class S theories associated with three punctured spheres, as is the case we are considering here. For theories associated with higher genus though, the two may deviate.}.

This holds for the pre-folding quiver. However, in \cite{Zafrir:2016wkk}, twisted reductions of $5d$ SCFTs were studied, which similarly to the case we are considering here, flow to class S theories, or theories related to them, when reduced without a twist. It was observed there that it is possible to formulate an expression for the Hall-Littlewood index associated with the twisted theory from the expression for the Hall-Littlewood index associated with the untwisted one. We shall generalize this idea to our case and use it to test the proposed magnetic quiver.

Let us begin with some preliminaries. First we remind the reader of the definition of the Hall-Littlewood index \cite{Gadde:2011uv}:

\be
I_{HL} = Tr_{HL} (-1)^F \tau^{2E-2R} \prod_i a^{f_i}_i ,
\ee
where $\tau$ is the fugacity associated with the superconformal algebra, and $a_i$ are fugacities associated with various flavor symmetries whose Cartan charges are given by $f_i$. Here $Tr_{HL}$ denotes trace over all operators obeying: $j_1=0$, $E-2R-r=0$, for $E$ the dimension of the operator, $j_1$ one of its highest weights under the $SO(4)$ rotation symmetry, $R$ its highest weight under the $SU(2)$ R-symmetry and $r$ its $U(1)$ R-symmetry charge.  

Next, consider a class S theory associated with the compactification of a $6d$ $(2,0)$ theory of type $G=A_{N-1}$ or $D_N$ on a Riemann sphere with punctures. Then the generic form of the Hall-Littlewood index was worked out in \cite{Gadde:2011uv,Gaiotto:2012uq,Gaiotto:2012xa,Lemos:2012ph} to be:

\be
I_{HL} = \mathcal{A}(\tau) \sum_{\lambda} \frac{\prod_{i}\mathcal{K}(\Lambda'(a_i)) \psi^G_{\lambda}(\Lambda(a_i))}{\psi^G_{\lambda}(\Lambda_{trivial})},
\ee 
where:

\begin{itemize} 

\item $\mathcal{A}(\tau)$ is a flavor fugacity independent normalization factor given by:

\be
\mathcal{A}(\tau) = (1-\tau^2)^n \prod_j (1-\tau^{2j}) .
\ee

Here $j$ runs over all the dimensions of the invariant polynomials of $G$ and $n$ is some number, which depends on $G$, see \cite{Gaiotto:2012uq,Lemos:2012ph,Chacaltana:2013oka}. 

\item The sum is over all the irreducible representations of the group $G$, which can be described by  partitions $\lambda$. For the $G=A_{N-1}$ case we have $\lambda=(\lambda_1, \lambda_2,...,\lambda_{N-1},0)$, with the sum going over all $\lambda_1\geq \lambda_2 \geq ... \geq \lambda_{N-1}\geq 0$. The case of $G=D_N$ is more involved. We first have $\lambda=(\lambda_1, \lambda_2,...,\lambda_{N-1},\lambda_{N})$ for $\lambda_1\geq \lambda_2 \geq ... \geq \lambda_{N}\geq 0$, which gives the representations of $SO(2N)$. This needs to be supplemented with partitions $\lambda$ where all $\lambda_i$ are half-integer, with $\lambda_N$ allowed to be negative. The product is over all the punctures.

\item $\mathcal{K}(\Lambda'(a_i))$ are fugacity dependent factors associated with each puncture. The exact expression for them can be found in \cite{Gaiotto:2012uq,Lemos:2012ph}.

\item $\psi^G_{\lambda}(\Lambda(a_i))$ are the Hall-Littlewood polynomials for the group $G$. They are given by:

\be
\psi^{A_{N-1}}_{\lambda}(\Lambda(a_i)) = \mathcal{N}_{\lambda} (\tau) \sum_{\sigma\subset S_N} x^{\lambda_1}_{\sigma(1)} ... x^{\lambda_N}_{\sigma(N)} \prod_{i<j} \frac{x_{\sigma(i)}- \tau^2 x_{\sigma(j)}}{x_{\sigma(i)}-x_{\sigma(j)}}
\ee

for $G=A_{N-1}$ and by:

\be
\psi^{D_{N}}_{\lambda}(\Lambda(a_i)) = \mathcal{N}_{\lambda} (\tau) \sum_{\sigma\subset S_N} \sum_{\stackrel{s_1,...,s_N=\pm 1}{\prod s_i = +1}} x^{s_1\lambda_1}_{\sigma(1)} ... x^{\lambda_N}_{s_N\sigma(N)} \prod_{i<j} \frac{x^{s_i}_{\sigma(i)}- \tau^2 x^{\pm s_j}_{\sigma(j)}}{x^{s_i}_{\sigma(i)}-x^{\pm s_j}_{\sigma(j)}}
\ee

for $G=D_N$. 

Here $\mathcal{N}_{\lambda} (\tau)$ is a normalization factor given by:

\be
\mathcal{N}^{-2}_{\lambda} (\tau) = \sum_{\stackrel{w\in W_G}{w \lambda = \lambda}} \tau^{2l(w)},
\ee
where $W_G$ is the Weyl group of $G$ and $l(w)$ denotes the length of the Weyl group element $w$.

\item $\Lambda(a_i)$ is a list of N elements whose exact form depends on the type of puncture. The procedure for determining it in the general case can be found in \cite{Gaiotto:2012uq,Lemos:2012ph}. $\Lambda_{trivial}$ describes the list for the trivial partition, which for our case of interest is given by $\Lambda_{trivial} = (\tau^{1-N},\tau^{3-N}, ... , \tau^{N-1})$ for $G=A_{N-1}$ and by $\Lambda_{trivial} = (1,\tau^2,\tau^4, ... , ,\tau^{2N-2})$ for $G=D_{N}$.

\end{itemize}

This gives the Hall-Littlewood index for the case of a punctured sphere. Now consider the case where two or more of the punctures are identical. We then have a discrete symmetry exchanging the identical punctures, and we can consider twisting by that symmetry when performing the reduction. \cite{Zafrir:2016wkk} considered $5d$ SCFTs that reduce to $A$ type class S theories with identical punctures and further considered their reduction with a twist that is related to the discrete symmetry that acts in $4d$ by exchanging the identical punctures. One of the observations made there is that we can use the expression for the Hall-Littlewood index of the $4d$ theory we get without the twist to formulate an expression for the Hall-Littlewood index of the $4d$ theory resulting from the compactification with the twist. In general, the Hall-Littlewood index of the twisted theory can be expressed as:

\be
I_{HL} = (1-\tau^2)^q \mathcal{A}(\tau) \sum_{\lambda} \mathcal{N}^p_{\lambda}\frac{\prod_{i}\mathcal{K}(\Lambda'(a_i)) \mathcal{N}^{-1}_{l_i\lambda} \psi^G_{l_i\lambda}(\Lambda(a_i))}{(\tau)\psi^G_{\lambda}(\Lambda_{trivial})},
\ee 
where $p$ is the number of punctures in the pre-twisted theory and $q$ is some number that depends on the specific twist involved, see \cite{Zafrir:2016wkk}. Here the sum runs over orbits of punctures under the twist symmetry and $l_i$ denotes the length of the $i$ orbit. We also use $l_i\lambda$ to mean the partition given by $(l_i\lambda_1,l_i\lambda_2,...,l_i\lambda_{N-1},l_i\lambda_{N})$. We can understand this expression as implying that we need to identify the $l$ punctures related by the discrete symmetry and replace them with one puncture of the same type, but associated with representations that is the $l$ symmetric product of the representations of $G$ we are summing over. One subtlety is that the normalization factor $\mathcal{N}_{\lambda}$, coming from each puncture should remain unidentified\footnote{We note that $\mathcal{N}_{\lambda}=\mathcal{N}_{l \lambda}$.}.

This expression was noted and tested for the case of $G=A$, but we can also apply it to our purposes where $G=D$. Recall that we wish to consider the compactification of a $5d$ SCFT, whose direct $4d$ reduction gives a $D$ type class S theory associated with a three punctured sphere with two identical punctures, with a $\mathbb{Z}_2$ twist acting as the exchange of the two identical punctures. The above observation then suggests that the expression for the Hall-Littlewood index of the $4d$ theory resulting from the twisted compactification can be expressed as:   

\be \label{IndTwistDN}
I_{HL} = \mathcal{A}'(\tau) \sum_{\lambda} \mathcal{N}_{\lambda}(\tau) \frac{\mathcal{K}(\Lambda'(a_1)) \mathcal{K}(\Lambda'(a_2)) \psi^{D_N}_{\lambda}(\Lambda(a_1)) \psi^{D_N}_{2\lambda}(\Lambda(a_2))}{\psi^{D_N}_{\lambda}(1,\tau^2,\tau^4, ... ,\tau^{2N-2})},
\ee
where we label the puncture invariant under the exchange by $1$. Here $\mathcal{A}'(\tau)$ is equal to:

\be
\mathcal{A}'(\tau) = (1-\tau^{2N})\prod^{N-1}_{j=1} (1-\tau^{4j}).
\ee

Before applying it to the case at hand, it is convenient to test this expression in a known example. For this we take the case of the rank $1$ MN $E_8$ theory. This theory can be realized by the compactification of the $5d$ SCFT UV completing the $5d$ gauge theory with an $SU(2)$ gauge group and seven doublet hypermultiplets. By considering $SU(2)$ as $USp(2)$, we can engineer this $5d$ SCFT using a brane system involving an O$5^-$ plane, which when compactified to $4d$ has a description as a D type class S theory \cite{Zafrir:2016jpu}. The specific theory one finds is the $D_4$ $(2,0)$ theory on a sphere with two maximal punctures and a minimal one, which indeed describes the rank $1$ MN $E_8$ theory \cite{Chacaltana:2011ze}. In particular, this description has two identical punctures so we can also consider the $\mathbb{Z}_2$ twisted compactification. This should lead to the rank $1$ MN $E_7$ theory\footnote{This can be understood as follows. The $\mathbb{Z}_2$ acts on the global symmetry by the exchange of the two independent $SO(8)$ subgroups of $SO(16)\subset E_8$. This is an inner automorphism of $E_8$ and so is just an holonomy in $E_8$ generating a mass deformation. By projecting the $E_8$ currents to the ones invariant under this exchange, one can see that this deformation preserves the $E_7$ subgroup of $E_8$, suggesting it should lead to the rank $1$ MN $E_7$ theory.} \cite{Bourget:2021xex}. Next we can employ \eqref{IndTwistDN} to compute the Hall-Littlewood index of the resulting theory, and we indeed find agreement with the known Hall-Littlewood index of the rank $1$ $E_7$ theory, at least to the order we computed.

We can then apply this expression to the case at hand. We consider the $5d$ SCFT UV completing the $5d$ gauge theory $SU(4)_0+1AS+8F$. By regarding it as an $SO(6)$ gauge theory with vector and spinor matter, we can engineer it using a brane system involving an O$5^-$ plane. When reduced to $4d$ then, we expect to get a D type class S theory, which in our case turns out to be $D_5$ on a sphere with three punctures corresponding to the partitions: $(8,2)$, $(8,2)$ and $(2^5)$. This describes a rank $3$ $4d$ $\mathcal{N}=2$ SCFT with $SU(4)\times SU(8)$ global symmetry which is the result of the direct reduction of this $5d$ SCFT. As advertized, it has two identical punctures implying the presence of a $\mathbb{Z}_2$ symmetry exchanging them. The latter appears to act on the global symmetry as charge conjugation so it appears to have the right properties to be the $\mathbb{Z}_2$ we are after. Additionally, we noted that the magnetic quiver resulting from folding has the right dimension to be that of the $4d$ $USp(4) \times USp(8)$ SCFT. As a final check, we can apply \eqref{IndTwistDN} to compute the Hall-Littlewood index. We find:

\be
I_{HL} = 1 + 46\tau^2 + 108\tau^3 + 1290\tau^4 + 4716\tau^5 + O(\tau^6) ,
\ee
where here we have unrefined with respect to the flavor fugacities to simplify the computation. This can then be compared against the one of the $USp(4) \times USp(8)$ SCFT, evaluated for instance using the class S description in \cite{Chacaltana:2011ze,Chacaltana:2013oka} finding perfect agreement\footnote{We thank Behzat Ergun for sharing with us a preliminary version of his Mathematica package to compute Schur and Hall-Littlewood indices of class-S theories.}.

\section{\texorpdfstring{Moduli space of $\mathcal{N}=3$ theories}{Moduli space of N=3 theories}}\label{app:N=3}

In the case of $\cN=3$ theories, it is possible to compute the stratification of the HB explicitly as the HB of either theories can be written as an orbifold:
\begin{align}
    \cH\left(\cT^{\cN=3}\right)=\C^4/(\Gamma\oplus\bar{\Gamma})
\end{align}
where $\bar{\phantom{\Gamma}}$ indicate complex conjugation and $\G$ should be identified with the irreducible action of the rank-2 complex reflection groups $G(3,1,2)$ and $G(4,1,2)$. Below we compute explicitly the fixed loci of this action which are naturally identified with the singular loci of the HB. We carry out the calculation explicitly for $G(3,1,2)$, the analysis in the $G(4,1,2)$ case is extremely similar and thus in this case we only report the results.

\subsection{\texorpdfstring{$G(3,1,2)$}{G(3,1,2)}}

Explicitly expanding the Hilbert series \eqref{HilbG312} we find that the HB chiral ring of the $\cN=3$ $G(3,1,2)$ theory is generated by eight elements:
\begin{align}
U,&\quad h=2\\
X, \bar{X}, &\quad h=3\\
Y,&\quad h=4\\
W, \bar{W}, &\quad h=5\\
Z, \bar{Z}, &\quad h=6
\end{align}
where $h$ indicates their scaling dimensions. These generators can be identified with the following invariant of the orbifold action:
\beq
\label{8invariants}
\begin{array}{c}
U=z_1 \bz_1+z_2\bz_2,\quad X=z_1^3+z_2^3,\quad \bar{X}=\bz_1^3+\bz_2^3,\\
Y=z_1z_2\bz_1\bz_2,\quad W=z^4_1 \bz_1+z^4_2\bz_2,\quad \bar{W}=\bz^4_1 z_1+\bz^4_2 z_2,\\
Z=z_1^6+z_2^6,\quad \bar{Z}=\bz_1^6+\bz_2^6.\\
\end{array}
\eeq
Here ${\bf z}=(z_1,z_2)$ are each a doublet of coordinates under the $SU(2)$ R-symmetry group that acts on the Higgs branch. This gives a total of four complex coordinates spanning $\C^4$. We also use the explicit matrix representation of $G(3,1,2)$ generated by:
\begin{align}
    M_1=\left(
    \begin{array}{cc}
    \exp(2i \pi/3)&0\\
    0 &1
    \end{array}
    \right),\qquad M_2=\left(
    \begin{array}{cc}
    0&1\\
    1&0
    \end{array}
    \right)
\end{align}
As a check, the Hilbert series for the variety parametrized by the eight invariants (\ref{8invariants}) can be computed and it indeed agrees with \eqref{HilbG312}.

To identify the HB stratification we should analyze the fixed locus of the $G(3,1,2)$ action. Given that we know the explicit action, this can be done straightforwardly by solving the following linear equation in the ${\bf z}$:
\beq\label{fixloc}
M {\bf z}={\bf z},\quad M\in G(3,1,2)
\eeq

Solutions of \eqref{fixloc} which lie on the same $G(3,1,2)$ orbit need to be identified as they provide equivalent characterization of the same connected locus. Alternatively, we could evaluate the invariant polynomials on the solutions and report only those which are inequivalent. We follow this latter approach which has the advantage of also directly providing an explicit algebraic form for each singular locus. 

The solutions of \eqref{fixloc} are 
\begin{itemize}
\item The point ${\bf z} = 0$ is left invariant by all of $G(3,1,2)$. This is the origin of the HB. 
\item The subspace $z_1=z_2$, and those related to it by $G(3,1,2)$ orbits. This corresponds to the normal variety
\begin{align}
\bar{\cH}_1:& \quad \frac{\C[U,X,\bar{X},Y,W,\bar{W},Z,\bar{Z}]}{\langle U^3-2 X \bar{X},4Y-U^2,2W-U X,2\bar{W}-U \bar{X},2Z - X^2,2\bar{Z} - \bar{X}^2\rangle}\\\nonumber
&\qquad \cong\frac{\C[U,X,\bar{X}]}{\langle U^3-2X \bar{X}\rangle} 
\end{align}
\item The subspace $z_2=0$, and those related to it by $G(3,1,2)$ orbits. This corresponds to the normal variety
\begin{align}
\bar{\cH}_2:& \quad \frac{\C[U,X,\bar{X},Y,W,\bar{W},Z,\bar{Z}]}{\langle U^3-X \bar{X},Y,W-U X,\bar{W}-U \bar{X},Z-X^2,\bar{Z}-\bar{X}^2\rangle}\\\nonumber
&\qquad \cong \frac{\C[U,X,\bar{X}]}{\langle U^3-X \bar{X}\rangle} \, . 
\end{align}
\item The generic point ${\bf z}$, which is left invariant by the trivial subgroup of $G(3,1,2)$, corresponds to the highest dimensional leaf of the HB. 
\end{itemize}

Given this analysis we conclude that the HB of the $\cN=3$ $G(3,1,2)$ theory has two bottom elementary slices topologically $A_2\cong \C^2/\Z_3$. To identify the subsequent elementary slice we can perform an analysis along the lines of \cite{Martone:2021ixp} to identify the theory supported on each singular stratum on the CB, use $\cN=3$ SUSY to then infer that the same theories should be supported on singular strata of the HB and then identify the last two elementary slices as the HBs of the latter. This analysis results in the following Hasse diagram:
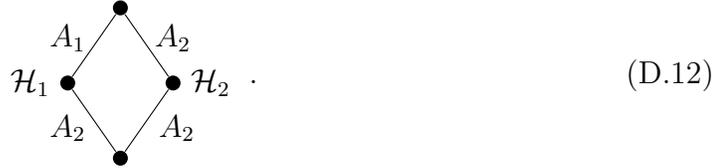
\begin{equation}
\label{hasseG312}
   \raisebox{-.5\height}{\begin{tikzpicture}
		\node[hasse] (a0) at (0,0) {};
		\node[hasse] (a1) at (.7,1) {};
		\node[hasse] (a2) at (-.7,1) {};
		\node[hasse] (a3) at (0,2) {};
		\draw (a0)--(a1)--(a3);
		\draw (a0)--(a2)--(a3);
		\node at (-0.7,0.4) {$A_{2}$};
		\node at (-0.7,1.6) {$A_{1}$};
		\node at (0.75,0.4) {$A_{2}$};
		\node at (0.7,1.6) {$A_{2}$};
		\node at (-1.2,1) {$\cH_1$};
		\node at (1.2,1) {$\cH_2$};
	\end{tikzpicture}} \;.
\end{equation}

We can also compute directly the transverse slices as follows. Let's begin with $\bar{\cH}_2$. In order to compute the transverse slice, one picks a generic point on $\bar{\cH}_2$, say $z_1 = a$. The transverse slice is then parametrized by the value of $z_2$. The invariants on this slice are then given by  
\begin{eqnarray}
& & U = |a|^2 + z_2 \bz_2 \qquad X = a^3 + z_2^3 \qquad \bar{X} = \bar{a}^3 + \bz_2^3 \qquad Y = |a|^2 z_2 \bz_2  \\ \nonumber & & W = |a|^2 a^3 + z^4_2 \bz_2 \qquad \bar{W} = |a|^2 \bar{a}^3 + \bz^4_2 z_2 \qquad Z = a^6 + z_2^6 \qquad \bar{Z} = \bar{a}^6 + \bz_2^6
\end{eqnarray}
We see that up to constant shifts depending on the $\bar{\cH}_2$ location, the transverse slice has the equation $U^3 - X \bar{X} = 0$, which is of type $A_2$. 

For $\bar{\cH}_1$, we proceed similarly. We first define $z_1=v+u$, $z_2=v-u$. A generic point on $\bar{\cH}_1$ is then given by $v$, while $u$ parametrizes the transverse slice. The invariants on this slice are then given by 
\begin{equation}
U = 2|v|^2+2|u|^2 \qquad X = 2v^3 + 6v u^2  \qquad \bar{X} = 2\bar{v}^3 + 6\bar{v} \bar{u}^2  
\end{equation}
with the other invariants being polynomial in these, up to shifts. Again we see that up to constant shifts depending on the $\bar{\cH}_1$ location, the equation of the transverse slice is $9 |v|^2 U^2 - X \bar{X} = 0$, which is of type $A_1$ for $v \neq 0$. This identifies the two transverse slices in (\ref{hasseG312}). 

It is interesting to compare this with the string theory construction of this SCFT given in \cite{Garcia-Etxebarria:2015wns}, as two D$3$-branes probing an S-fold singularity. The Higgs branch can be described by the location of the two D$3$-branes on the transverse $\mathbb{C}^2/G(3,1,2)$ subspace, which we can identify with $z_1$ and $z_2$. The point $z_1=z_2=0$ has both D$3$-branes sitting at the singularity, which corresponds to the SCFT point. We can consider two interesting directions. One is to pull out one of the D$3$-branes while leaving the other in the singular point. This describes the subset $z_2=0$. The second is to keep the two D$3$-branes coincident, but pull both out of the singular point. This describes the subset $z_1=z_2$.

We note that on the first subspace we get the rank $1$ $\mathcal{N}=3$ SCFT with Higgs branch $\mathbb{C}^2/\mathbb{Z}_3$, corresponding to the $A_2$ slice. However, on the second subspace we get the $SU(2)$ $\mathcal{N}=4$ theory, whose Higgs branch is $\mathbb{C}^2/\mathbb{Z}_2$, corresponding to the $A_1$ slice. Thus, we see that our results are consistent with the brane realization of this $4d$ SCFT.

\subsection{\texorpdfstring{$G(4,1,2)$}{G(4,1,2)}}

Performing a similar analysis by expanding the Hilbert series \eqref{HilbG412} we find that the HB chiral ring of the $\cN=3$ $G(4,1,2)$ theory is again generated by eight elements, but with slightly different scaling dimensions:
\begin{align}
U,&\quad h=2\\
X_1, \bar{X}_1, X_0,&\quad h=4\\
Y, \bar{Y},&\quad h=6\\
Z, \bar{Z},&\quad h=8
\end{align}
with 
\beq
\begin{array}{c}
U=z_1 \bz_1+z_2\bz_2, \\ 
 X_1=z_1^4 + z_2^4 ,\quad \bar{X}_1=\bz_1^4 + \bz_2^4, \quad X_0=z_1 \bz_1 z_2 \bz_2 \\
 Y = z_1^5 \bz_1 +  z_2^5 \bz_2 , \quad \bar{Y} = \bz_1^5 z_1 +  \bz^5_2 z_2 \\
 Z = z_1^8 + z_2^8 , \quad \bar{Z} = \bz_1^8 + \bz_2^8 
\end{array}
\eeq
The analysis of the singular locus proceeds as before and we find two interesting subsets again corresponding to the $z_1=z_2$ and $z_2=0$ cases, which we denote as $\bar{\cH}_1$ and $\bar{\cH}_2$ respectively. These are given by:
\begin{align}
\bar{\cH}_1:& \quad \frac{\C[U,X_0,X_1,\bar{X}_1,Y,\bar{Y},Z,\bar{Z}]}{\langle U^4-4X_1 \bar{X}_1,4X_0-U^2, 2Y-UX_1, 2\bar{Y}-U\bar{X}_1, 2Z-X_1^2,2\bar{Z}-\bar{X}_1^2\rangle} \\\nonumber & \qquad  \cong\frac{\C[U,X_1,\bar{X}_1]}{\langle U^4-4X_1 \bar{X}_1\rangle}\\
\bar{\cH}_2:& \quad \frac{\C[U,X_0,X_1,\bar{X}_1,Y,\bar{Y},Z,\bar{Z}]}{\langle U^4-X_1\bar{X}_1,X_0, Y-UX_1, \bar{Y}-U\bar{X}_1, Z-X_1^2,\bar{Z}-\bar{X}_1^2\rangle} \\\nonumber & \qquad  \cong\frac{\C[U,X_1,\bar{X}_1]}{\langle U^4-X_1 \bar{X}_1\rangle}
\end{align}
thus we conclude now that the HB of the $\cN=3$ $G(4,1,2)$ theory has two bottom elementary slices topologically $A_3\cong \C^2/\Z_4$. The rest can be identified as described above with the final Hasse diagram:
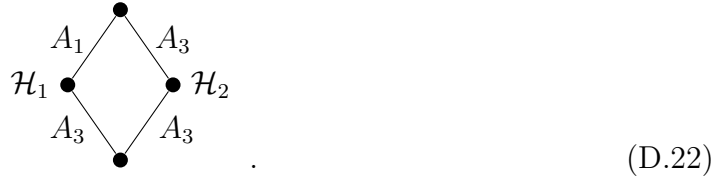
\begin{equation}
    \begin{tikzpicture}
		\node[hasse] (a0) at (0,0) {};
		\node[hasse] (a1) at (.7,1) {};
		\node[hasse] (a2) at (-.7,1) {};
		\node[hasse] (a3) at (0,2) {};
		\draw (a0)--(a1)--(a3);
		\draw (a0)--(a2)--(a3);
		\node at (-0.7,0.4) {$A_{3}$};
		\node at (-0.7,1.6) {$A_{1}$};
		\node at (0.75,0.4) {$A_{3}$};
		\node at (0.7,1.6) {$A_{3}$};
		\node at (-1.2,1) {$\cH_1$};
		\node at (1.2,1) {$\cH_2$};
	\end{tikzpicture}\;.
\end{equation}

We can again compare this with the string theory construction of this SCFT given in \cite{Garcia-Etxebarria:2015wns}, as two D$3$-branes probing an S-fold singularity. The only difference here is that the S-fold now has $\mathbb{Z}_4$ quotient instead of $\mathbb{Z}_3$. We still have the same two subspaces, but now on the $z_2=0$ subspace we have the rank $1$ $\mathcal{N}=3$ SCFT with Higgs branch $\mathbb{C}^2/\mathbb{Z}_4$. This gives an $A_3$ slice instead of the $A_2$ slice, again in agreement with the Hasse diagram we observe.

\providecommand{\href}[2]{#2}\begingroup\raggedright\endgroup


\begin{thebibliography}{10}

\bibitem{Dolan:2002zh}
F.~A. Dolan and H.~Osborn, {\it {On short and semi-short representations for
  four-dimensional superconformal symmetry}},  {\em Annals Phys.} {\bf 307}
  (2003) 41--89, [\href{http://arxiv.org/abs/hep-th/0209056}{{\tt
  hep-th/0209056}}].

\bibitem{Cordova:2016emh}
C.~Cordova, T.~T. Dumitrescu, and K.~Intriligator, {\it {Multiplets of
  Superconformal Symmetry in Diverse Dimensions}},
  \href{http://arxiv.org/abs/1612.00809}{{\tt arXiv:1612.00809}}.

\bibitem{Manenti:2019jds}
A.~Manenti, {\it {Differential operators for superconformal correlation
  functions}},  \href{http://arxiv.org/abs/1910.12869}{{\tt arXiv:1910.12869}}.

\bibitem{Buican:2014qla}
M.~Buican, T.~Nishinaka, and C.~Papageorgakis, {\it {Constraints on chiral
  operators in $ \mathcal{N}=2 $ SCFTs}},  {\em JHEP} {\bf 12} (2014) 095,
  [\href{http://arxiv.org/abs/1407.2835}{{\tt arXiv:1407.2835}}].

\bibitem{Caorsi:2018zsq}
M.~Caorsi and S.~Cecotti, {\it {Geometric classification of 4d $\mathcal{N}=2$
  SCFTs}},  {\em JHEP} {\bf 07} (2018) 138,
  [\href{http://arxiv.org/abs/1801.04542}{{\tt arXiv:1801.04542}}].

\bibitem{Argyres:2018urp}
P.~C. Argyres and M.~Martone, {\it {Scaling dimensions of Coulomb branch
  operators of 4d N=2 superconformal field theories}},
  \href{http://arxiv.org/abs/1801.06554}{{\tt arXiv:1801.06554}}.

\bibitem{Argyres:2015ffa}
P.~Argyres, M.~Lotito, Y.~L{\"u}, and M.~Martone, {\it {Geometric constraints
  on the space of $ \mathcal{N} $ = 2 SCFTs. Part I: physical constraints on
  relevant deformations}},  {\em JHEP} {\bf 02} (2018) 001,
  [\href{http://arxiv.org/abs/1505.04814}{{\tt arXiv:1505.04814}}].

\bibitem{Argyres:2015gha}
P.~C. Argyres, M.~Lotito, Y.~L{\"u}, and M.~Martone, {\it {Geometric
  constraints on the space of $ \mathcal{N} $ = 2 SCFTs. Part II: construction
  of special Kahler geometries and RG flows}},  {\em JHEP} {\bf 02} (2018) 002,
  [\href{http://arxiv.org/abs/1601.00011}{{\tt arXiv:1601.00011}}].

\bibitem{Argyres:2016xua}
P.~C. Argyres, M.~Lotito, Y.~L{\"u}, and M.~Martone, {\it {Expanding the
  landscape of $ \mathcal{N} $ = 2 rank 1 SCFTs}},  {\em JHEP} {\bf 05} (2016)
  088, [\href{http://arxiv.org/abs/1602.02764}{{\tt arXiv:1602.02764}}].

\bibitem{Argyres:2016xmc}
P.~Argyres, M.~Lotito, Y.~L{\"u}, and M.~Martone, {\it {Geometric constraints
  on the space of $ \mathcal{N}$ = 2 SCFTs. Part III: enhanced Coulomb branches
  and central charges}},  {\em JHEP} {\bf 02} (2018) 003,
  [\href{http://arxiv.org/abs/1609.04404}{{\tt arXiv:1609.04404}}].

\bibitem{Martone:2020nsy}
M.~Martone, {\it {Towards the classification of rank-$r$ $\mathcal{N}=2$ SCFTs.
  Part I: twisted partition function and central charge formulae}},
  \href{http://arxiv.org/abs/2006.16255}{{\tt arXiv:2006.16255}}.

\bibitem{Argyres:2020wmq}
P.~C. Argyres and M.~Martone, {\it {Towards a classification of rank $r$
  $\mathcal{N}=2$ SCFTs Part II: special Kahler stratification of the Coulomb
  branch}},  \href{http://arxiv.org/abs/2007.00012}{{\tt arXiv:2007.00012}}.

\bibitem{Shimizu:2017kzs}
H.~Shimizu, Y.~Tachikawa, and G.~Zafrir, {\it {Anomaly matching on the Higgs
  branch}},  {\em JHEP} {\bf 12} (2017) 127,
  [\href{http://arxiv.org/abs/1703.01013}{{\tt arXiv:1703.01013}}].

\bibitem{Beem:2013sza}
C.~Beem, M.~Lemos, P.~Liendo, W.~Peelaers, L.~Rastelli, and B.~C. van Rees,
  {\it {Infinite Chiral Symmetry in Four Dimensions}},  {\em Commun. Math.
  Phys.} {\bf 336} (2015), no.~3 1359--1433,
  [\href{http://arxiv.org/abs/1312.5344}{{\tt arXiv:1312.5344}}].

\bibitem{Lemos:2015orc}
M.~Lemos and P.~Liendo, {\it {$\mathcal{N}=2$ central charge bounds from $2d$
  chiral algebras}},  {\em JHEP} {\bf 04} (2016) 004,
  [\href{http://arxiv.org/abs/1511.07449}{{\tt arXiv:1511.07449}}].

\bibitem{Bourget:2020asf}
A.~Bourget, J.~F. Grimminger, A.~Hanany, M.~Sperling, G.~Zafrir, and Z.~Zhong,
  {\it {Magnetic quivers for rank 1 theories}},
  \href{http://arxiv.org/abs/2006.16994}{{\tt arXiv:2006.16994}}.

\bibitem{Martone:2021ixp}
M.~Martone, {\it {Testing our understanding of SCFTs: a catalogue of rank-2
  $\mathcal{N}$=2 theories in four dimensions}},
  \href{http://arxiv.org/abs/2102.02443}{{\tt arXiv:2102.02443}}.

\bibitem{Bourget:2020mez}
A.~Bourget, S.~Giacomelli, J.~F. Grimminger, A.~Hanany, M.~Sperling, and
  Z.~Zhong, {\it {S-fold magnetic quivers}},  {\em JHEP} {\bf 02} (2021) 054,
  [\href{http://arxiv.org/abs/2010.05889}{{\tt arXiv:2010.05889}}].

\bibitem{vanBeest:2021xyt}
M.~van Beest and S.~Giacomelli, {\it {Connecting 5d Higgs Branches via
  Fayet-Iliopoulos Deformations}},  \href{http://arxiv.org/abs/2110.02872}{{\tt
  arXiv:2110.02872}}.

\bibitem{Martone:2021drm}
M.~Martone and G.~Zafrir, {\it {On the compactification of 5d theories to 4d}},
   {\em JHEP} {\bf 08} (2021) 017, [\href{http://arxiv.org/abs/2106.00686}{{\tt
  arXiv:2106.00686}}].

\bibitem{Cremonesi:2014xha}
S.~Cremonesi, G.~Ferlito, A.~Hanany, and N.~Mekareeya, {\it {Coulomb Branch and
  The Moduli Space of Instantons}},  {\em JHEP} {\bf 12} (2014) 103,
  [\href{http://arxiv.org/abs/1408.6835}{{\tt arXiv:1408.6835}}].

\bibitem{Nakajima:2019olw}
H.~Nakajima and A.~Weekes, {\it {Coulomb branches of quiver gauge theories with
  symmetrizers}},  \href{http://arxiv.org/abs/1907.06552}{{\tt
  arXiv:1907.06552}}.

\bibitem{Bourget:2021xex}
A.~Bourget, J.~F. Grimminger, A.~Hanany, R.~Kalveks, M.~Sperling, and Z.~Zhong,
  {\it {Folding Orthosymplectic Quivers}},
  \href{http://arxiv.org/abs/2107.00754}{{\tt arXiv:2107.00754}}.

\bibitem{Bourget:2020gzi}
A.~Bourget, J.~F. Grimminger, A.~Hanany, M.~Sperling, and Z.~Zhong, {\it
  {Magnetic Quivers from Brane Webs with O5 Planes}},  {\em JHEP} {\bf 07}
  (2020) 204, [\href{http://arxiv.org/abs/2004.04082}{{\tt arXiv:2004.04082}}].

\bibitem{Bourget:2020xdz}
A.~Bourget, J.~F. Grimminger, A.~Hanany, R.~Kalveks, M.~Sperling, and Z.~Zhong,
  {\it {Magnetic Lattices for Orthosymplectic Quivers}},  {\em JHEP} {\bf 12}
  (2020) 092, [\href{http://arxiv.org/abs/2007.04667}{{\tt arXiv:2007.04667}}].

\bibitem{Bourget:2019rtl}
A.~Bourget, S.~Cabrera, J.~F. Grimminger, A.~Hanany, and Z.~Zhong, {\it {Brane
  Webs and Magnetic Quivers for SQCD}},  {\em JHEP} {\bf 03} (2020) 176,
  [\href{http://arxiv.org/abs/1909.00667}{{\tt arXiv:1909.00667}}].

\bibitem{Beem:2017ooy}
C.~Beem and L.~Rastelli, {\it {Vertex operator algebras, Higgs branches, and
  modular differential equations}},  {\em JHEP} {\bf 08} (2018) 114,
  [\href{http://arxiv.org/abs/1707.07679}{{\tt arXiv:1707.07679}}].

\bibitem{Cremonesi:2015lsa}
S.~Cremonesi, G.~Ferlito, A.~Hanany, and N.~Mekareeya, {\it {Instanton
  Operators and the Higgs Branch at Infinite Coupling}},  {\em JHEP} {\bf 04}
  (2017) 042, [\href{http://arxiv.org/abs/1505.06302}{{\tt arXiv:1505.06302}}].

\bibitem{Cabrera:2018jxt}
S.~Cabrera, A.~Hanany, and F.~Yagi, {\it {Tropical Geometry and Five
  Dimensional Higgs Branches at Infinite Coupling}},  {\em JHEP} {\bf 01}
  (2019) 068, [\href{http://arxiv.org/abs/1810.01379}{{\tt arXiv:1810.01379}}].

\bibitem{Akhond:2020vhc}
M.~Akhond, F.~Carta, S.~Dwivedi, H.~Hayashi, S.-S. Kim, and F.~Yagi, {\it
  {Five-brane webs, Higgs branches and unitary/orthosymplectic magnetic
  quivers}},  {\em JHEP} {\bf 12} (2020) 164,
  [\href{http://arxiv.org/abs/2008.01027}{{\tt arXiv:2008.01027}}].

\bibitem{Benini:2009gi}
F.~Benini, S.~Benvenuti, and Y.~Tachikawa, {\it {Webs of five-branes and N=2
  superconformal field theories}},  {\em JHEP} {\bf 09} (2009) 052,
  [\href{http://arxiv.org/abs/0906.0359}{{\tt arXiv:0906.0359}}].

\bibitem{vanBeest:2020kou}
M.~van Beest, A.~Bourget, J.~Eckhard, and S.~Schafer-Nameki, {\it {(Symplectic)
  Leaves and (5d Higgs) Branches in the Poly(go)nesian Tropical Rain Forest}},
  {\em JHEP} {\bf 11} (2020) 124, [\href{http://arxiv.org/abs/2008.05577}{{\tt
  arXiv:2008.05577}}].

\bibitem{vanBeest:2020civ}
M.~Van~Beest, A.~Bourget, J.~Eckhard, and S.~Sch\"afer-Nameki, {\it {(5d
  RG-flow) Trees in the Tropical Rain Forest}},  {\em JHEP} {\bf 03} (2021)
  241, [\href{http://arxiv.org/abs/2011.07033}{{\tt arXiv:2011.07033}}].

\bibitem{Benini:2010uu}
F.~Benini, Y.~Tachikawa, and D.~Xie, {\it {Mirrors of 3d Sicilian theories}},
  {\em JHEP} {\bf 09} (2010) 063, [\href{http://arxiv.org/abs/1007.0992}{{\tt
  arXiv:1007.0992}}].

\bibitem{Beratto:2020wmn}
E.~Beratto, S.~Giacomelli, N.~Mekareeya, and M.~Sacchi, {\it {3d mirrors of the
  circle reduction of twisted A$_{2N}$ theories of class S}},  {\em JHEP} {\bf
  09} (2020) 161, [\href{http://arxiv.org/abs/2007.05019}{{\tt
  arXiv:2007.05019}}].

\bibitem{Chacaltana:2012ch}
O.~Chacaltana, J.~Distler, and Y.~Tachikawa, {\it {Gaiotto duality for the
  twisted A$_{2N?1}$ series}},  {\em JHEP} {\bf 05} (2015) 075,
  [\href{http://arxiv.org/abs/1212.3952}{{\tt arXiv:1212.3952}}].

\bibitem{Chacaltana:2013oka}
O.~Chacaltana, J.~Distler, and A.~Trimm, {\it {Tinkertoys for the Twisted
  D-Series}},  {\em JHEP} {\bf 04} (2015) 173,
  [\href{http://arxiv.org/abs/1309.2299}{{\tt arXiv:1309.2299}}].

\bibitem{Beem:2019snk}
C.~Beem, C.~Meneghelli, W.~Peelaers, and L.~Rastelli, {\it {VOAs and rank-two
  instanton SCFTs}},  \href{http://arxiv.org/abs/1907.08629}{{\tt
  arXiv:1907.08629}}.

\bibitem{Garcia-Etxebarria:2015wns}
I.~Garc{\'i}a-Etxebarria and D.~Regalado, {\it {$ \mathcal{N}=3 $ four
  dimensional field theories}},  {\em JHEP} {\bf 03} (2016) 083,
  [\href{http://arxiv.org/abs/1512.06434}{{\tt arXiv:1512.06434}}].

\bibitem{Aharony:2016kai}
O.~Aharony and Y.~Tachikawa, {\it {S-folds and 4d N=3 superconformal field
  theories}},  {\em JHEP} {\bf 06} (2016) 044,
  [\href{http://arxiv.org/abs/1602.08638}{{\tt arXiv:1602.08638}}].

\bibitem{Apruzzi:2020pmv}
F.~Apruzzi, S.~Giacomelli, and S.~Sch\"afer-Nameki, {\it {4d $\mathcal{N}=2$
  S-folds}},  {\em Phys. Rev. D} {\bf 101} (2020), no.~10 106008,
  [\href{http://arxiv.org/abs/2001.00533}{{\tt arXiv:2001.00533}}].

\bibitem{Giacomelli:2020jel}
S.~Giacomelli, C.~Meneghelli, and W.~Peelaers, {\it {New N=2 superconformal
  field theories from S-folds}},  \href{http://arxiv.org/abs/2007.00647}{{\tt
  arXiv:2007.00647}}.

\bibitem{Heckman:2020svr}
J.~J. Heckman, C.~Lawrie, T.~B. Rochais, H.~Y. Zhang, and G.~Zoccarato, {\it
  {S-folds, String Junctions, and 4D $\mathcal{N} = 2$ SCFTs}},
  \href{http://arxiv.org/abs/2009.10090}{{\tt arXiv:2009.10090}}.

\bibitem{Giacomelli:2020gee}
S.~Giacomelli, M.~Martone, Y.~Tachikawa, and G.~Zafrir, {\it {More on
  $\mathcal{N} =2$ S-folds}},  {\em JHEP} {\bf 01} (2021) 054,
  [\href{http://arxiv.org/abs/2010.03943}{{\tt arXiv:2010.03943}}].

\bibitem{Song:2017oew}
J.~Song, D.~Xie, and W.~Yan, {\it {Vertex operator algebras of Argyres-Douglas
  theories from M5-branes}},  {\em JHEP} {\bf 12} (2017) 123,
  [\href{http://arxiv.org/abs/1706.01607}{{\tt arXiv:1706.01607}}].

\bibitem{Dedushenko:2019mnd}
M.~Dedushenko and Y.~Wang, {\it {4d/2d $\rightarrow $ 3d/1d: A song of
  protected operator algebras}},  \href{http://arxiv.org/abs/1912.01006}{{\tt
  arXiv:1912.01006}}.

\bibitem{Closset:2020scj}
C.~Closset, S.~Schafer-Nameki, and Y.-N. Wang, {\it {Coulomb and Higgs Branches
  from Canonical Singularities: Part 0}},  {\em JHEP} {\bf 02} (2021) 003,
  [\href{http://arxiv.org/abs/2007.15600}{{\tt arXiv:2007.15600}}].

\bibitem{Giacomelli:2020ryy}
S.~Giacomelli, N.~Mekareeya, and M.~Sacchi, {\it {New aspects of
  Argyres--Douglas theories and their dimensional reduction}},  {\em JHEP} {\bf
  03} (2021) 242, [\href{http://arxiv.org/abs/2012.12852}{{\tt
  arXiv:2012.12852}}].

\bibitem{Xie:2021ewm}
D.~Xie, {\it {3d mirror for Argyres-Douglas theories}},
  \href{http://arxiv.org/abs/2107.05258}{{\tt arXiv:2107.05258}}.

\bibitem{arakawa2018joseph}
T.~Arakawa and A.~Moreau, {\it Joseph ideals and lisse minimal-algebras},  {\em
  Journal of the Institute of Mathematics of Jussieu} {\bf 17} (2018), no.~2
  397--417.

\bibitem{Beem:2019tfp}
C.~Beem, C.~Meneghelli, and L.~Rastelli, {\it {Free Field Realizations from the
  Higgs Branch}},  {\em JHEP} {\bf 09} (2019) 058,
  [\href{http://arxiv.org/abs/1903.07624}{{\tt arXiv:1903.07624}}].

\bibitem{Gledhill:2021cbe}
K.~Gledhill and A.~Hanany, {\it {Coulomb Branch Global Symmetry and Quiver
  Addition}},  \href{http://arxiv.org/abs/2109.07237}{{\tt arXiv:2109.07237}}.

\bibitem{Bergman:2015dpa}
O.~Bergman and G.~Zafrir, {\it {5d fixed points from brane webs and
  O7-planes}},  {\em JHEP} {\bf 12} (2015) 163,
  [\href{http://arxiv.org/abs/1507.03860}{{\tt arXiv:1507.03860}}].

\bibitem{Zafrir:2015ftn}
G.~Zafrir, {\it {Brane webs and $O5$-planes}},  {\em JHEP} {\bf 03} (2016) 109,
  [\href{http://arxiv.org/abs/1512.08114}{{\tt arXiv:1512.08114}}].

\bibitem{Zafrir:2016jpu}
G.~Zafrir, {\it {Brane webs in the presence of an O5$^{-}$-plane and 4d class S
  theories of type D}},  {\em JHEP} {\bf 07} (2016) 035,
  [\href{http://arxiv.org/abs/1602.00130}{{\tt arXiv:1602.00130}}].

\bibitem{Argyres:2019ngz}
P.~C. Argyres, A.~Bourget, and M.~Martone, {\it {Classification of all
  $\mathcal{N}\geq 3$ moduli space orbifold geometries at rank 2}},
  \href{http://arxiv.org/abs/1904.10969}{{\tt arXiv:1904.10969}}.

\bibitem{Cabrera:2018ann}
S.~Cabrera and A.~Hanany, {\it {Quiver Subtractions}},  {\em JHEP} {\bf 09}
  (2018) 008, [\href{http://arxiv.org/abs/1803.11205}{{\tt arXiv:1803.11205}}].

\bibitem{Bourget:2019aer}
A.~Bourget, S.~Cabrera, J.~F. Grimminger, A.~Hanany, M.~Sperling, A.~Zajac, and
  Z.~Zhong, {\it {The Higgs mechanism --- Hasse diagrams for symplectic
  singularities}},  {\em JHEP} {\bf 01} (2020) 157,
  [\href{http://arxiv.org/abs/1908.04245}{{\tt arXiv:1908.04245}}].

\bibitem{Grimminger:2020dmg}
J.~F. Grimminger and A.~Hanany, {\it {Hasse diagrams for 3d $ \mathcal{N} $ = 4
  quiver gauge theories \textemdash{} Inversion and the full moduli space}},
  {\em JHEP} {\bf 09} (2020) 159, [\href{http://arxiv.org/abs/2004.01675}{{\tt
  arXiv:2004.01675}}].

\bibitem{Bourget:2021siw}
A.~Bourget, J.~F. Grimminger, A.~Hanany, M.~Sperling, and Z.~Zhong, {\it
  {Branes, Quivers, and the Affine Grassmannian}},
  \href{http://arxiv.org/abs/2102.06190}{{\tt arXiv:2102.06190}}.

\bibitem{2003math......5095M}
A.~{Malkin}, V.~{Ostrik}, and M.~{Vybornov}, {\it {The minimal degeneration
  singularities in the affine Grassmannians}},  {\em arXiv Mathematics
  e-prints} (May, 2003) math/0305095,
  [\href{http://arxiv.org/abs/math/0305095}{{\tt math/0305095}}].

\bibitem{InstantonFuture}
A.~Bourget, J.~Grimminger, A.~Hanany, and Z.~Zhong, {\it On the hasse diagram
  of the moduli space of instantons}, .

\bibitem{Gaiotto:2012uq}
D.~Gaiotto and S.~S. Razamat, {\it {Exceptional Indices}},  {\em JHEP} {\bf 05}
  (2012) 145, [\href{http://arxiv.org/abs/1203.5517}{{\tt arXiv:1203.5517}}].

\bibitem{Hanany:2012dm}
A.~Hanany, N.~Mekareeya, and S.~S. Razamat, {\it {Hilbert Series for Moduli
  Spaces of Two Instantons}},  {\em JHEP} {\bf 01} (2013) 070,
  [\href{http://arxiv.org/abs/1205.4741}{{\tt arXiv:1205.4741}}].

\bibitem{Keller:2012da}
C.~A. Keller and J.~Song, {\it {Counting Exceptional Instantons}},  {\em JHEP}
  {\bf 07} (2012) 085, [\href{http://arxiv.org/abs/1205.4722}{{\tt
  arXiv:1205.4722}}].

\bibitem{Zafrir:2015rga}
G.~Zafrir, {\it {Brane webs, $5d$ gauge theories and $6d$ $\mathcal{N}=(1,0)$
  SCFT's}},  {\em JHEP} {\bf 12} (2015) 157,
  [\href{http://arxiv.org/abs/1509.02016}{{\tt arXiv:1509.02016}}].

\bibitem{Hayashi:2018bkd}
H.~Hayashi, S.-S. Kim, K.~Lee, and F.~Yagi, {\it {5-brane webs for 5d $
  \mathcal{N} $ = 1 G$_{2}$ gauge theories}},  {\em JHEP} {\bf 03} (2018) 125,
  [\href{http://arxiv.org/abs/1801.03916}{{\tt arXiv:1801.03916}}].

\bibitem{Zafrir:2016wkk}
G.~Zafrir, {\it {Compactifications of 5d SCFTs with a twist}},  {\em JHEP} {\bf
  01} (2017) 097, [\href{http://arxiv.org/abs/1605.08337}{{\tt
  arXiv:1605.08337}}].

\bibitem{Gadde:2011uv}
A.~Gadde, L.~Rastelli, S.~S. Razamat, and W.~Yan, {\it {Gauge Theories and
  Macdonald Polynomials}},  {\em Commun. Math. Phys.} {\bf 319} (2013)
  147--193, [\href{http://arxiv.org/abs/1110.3740}{{\tt arXiv:1110.3740}}].

\bibitem{Gaiotto:2012xa}
D.~Gaiotto, L.~Rastelli, and S.~S. Razamat, {\it {Bootstrapping the
  superconformal index with surface defects}},  {\em JHEP} {\bf 01} (2013) 022,
  [\href{http://arxiv.org/abs/1207.3577}{{\tt arXiv:1207.3577}}].

\bibitem{Lemos:2012ph}
M.~Lemos, W.~Peelaers, and L.~Rastelli, {\it {The superconformal index of class
  $S$ theories of type $D$}},  {\em JHEP} {\bf 05} (2014) 120,
  [\href{http://arxiv.org/abs/1212.1271}{{\tt arXiv:1212.1271}}].

\bibitem{Chacaltana:2011ze}
O.~Chacaltana and J.~Distler, {\it {Tinkertoys for the $D_N$ series}},  {\em
  JHEP} {\bf 02} (2013) 110, [\href{http://arxiv.org/abs/1106.5410}{{\tt
  arXiv:1106.5410}}].

\end{thebibliography}
\end{document}